\documentclass[journal]{IEEEtran}%
\usepackage{amsfonts}
\usepackage{amsmath}
\usepackage{amssymb}
\usepackage{amsthm}
\usepackage{graphicx}%
\usepackage{braket}
\usepackage{hyperref}
\usepackage{braket}
\usepackage{mathtools}
\usepackage{units,bbm}
\usepackage{verbatim}

\providecommand{\U}[1]{\protect\rule{.1in}{.1in}}
\newtheorem{theorem}{Theorem}

\newtheorem{lemma}[theorem]{Lemma}

\newtheorem{remark}[theorem]{Remark}

\def\W{\mathsf{W}}
\def\N{\mathcal{N}}
\def\tr{{\rm Tr}}
\def\arikan{Ar\i{}kan{}}

\let\originalleft\left
\let\originalright\right
\def\left#1{\mathopen{}\originalleft#1}
\def\right#1{\originalright#1\mathclose{}}

\begin{document}

\title{Polar codes for private and quantum communication over arbitrary channels}
\author{Joseph M.~Renes and Mark M.~Wilde\thanks{JMR was supported by the Swiss National Science
    Foundation (SNF) through the National Centre of Competence in
    Research ``Quantum Science and Technology'' and project
    No. 200020-135048 and PP00P2-128455, by the European Research
    Council (ERC) via grant No. 258932.  MMW\ acknowledges support from the Centre de Recherches
Math\'{e}matiques at the University of Montreal.

Joseph M.\ Renes is with the Institute for Theoretical Physics, ETH Zurich, Z\"urich, Switzerland. Mark M.~Wilde is with the School of Computer Science, McGill University, Montr\'eal, Qu\'ebec, Canada.

This work was presented in part at the 2012 International Symposium on Information Theory in
Cambridge, Massachusetts, USA and the
2012 International Symposium on Information Theory and its Applications in Honolulu, Hawaii, USA.}}
\maketitle
\begin{abstract}
We construct new polar coding schemes for the transmission of quantum or private classical information over arbitrary quantum channels. In the former case, our coding scheme achieves the symmetric coherent information and in the latter the symmetric private information. Both schemes are built from a polar coding construction capable of  transmitting classical information over a quantum channel [Wilde and Guha, IEEE Transactions on Information Theory, in press]. Appropriately merging two such classical-quantum schemes, one for transmitting ``amplitude'' information and the other for transmitting ``phase,'' leads to the new private and quantum coding schemes, similar to the construction for Pauli and erasure channels in [Renes, Dupuis, and Renner, Physical Review Letters 109, 050504 (2012)]. The encoding is entirely similar to the classical case, and thus efficient. The decoding can also be performed by successive cancellation, as in the classical case, but no efficient successive cancellation scheme is yet known for arbitrary quantum channels. An efficient code construction is unfortunately still unknown. Generally, our two coding schemes require entanglement or secret-key assistance, respectively, but we extend two known conditions under which the needed assistance rate vanishes. Finally, although our results are formulated for qubit channels, we show how the scheme can be extended to multiple qubits. This then demonstrates a near-explicit coding method for realizing one of the most striking phenomena in
quantum information theory: the \textit{superactivation effect}, whereby two
quantum channels which individually have zero quantum capacity can have a
non-zero quantum capacity when used together.

\end{abstract}

\IEEEPARstart{P}{olar} coding is a promising code construction for transmitting classical
information over classical channels~\cite{arikan_channel_2009}. \arikan\ proved that polar codes
achieve the symmetric capacity of any classical channel, with an
encoding and decoding complexity that is $O\left(  N\log N\right)  $ where $N$
is the number of channel uses. These codes exploit the channel polarization
effect whereby a particular recursive encoding induces a set of virtual
channels, such that some of the virtual channels are perfect for data
transmission while the others are useless for this task. The fraction
containing perfect virtual channels is equal to the channel's symmetric
capacity.\footnote{The symmetric capacity is equal to the channel's input-output mutual information, evaulated for a uniformly random input.}

In this paper, we offer new polar coding schemes for transmitting quantum information or for privately transmitting classical information. Both are strongly based on
ideas of Renes and Boileau \cite{renes_physical_2008}, who showed that quantum or private coding protocols can be constructed from two different protocols
that protect classical information encoded into complementary variables. In
particular, a protocol for reliably transmitting quantum data can be built
from a protocol that reliably recovers classical information encoded into an
\textquotedblleft amplitude\textquotedblright\ variable and a protocol that
reliably recovers \textquotedblleft phase\textquotedblright\ information with
the assistance of quantum side information. The quantum coding scheme uses the decoders of both of these tasks, while the private coding scheme needs only the decoder of the amplitude variable and uses the fact that the phase could have been decoded in order to ensure security of the data via an entropic uncertainty relation   (see~\cite{renes_conjectured_2009, boileau_optimal_2009,renes_duality_2011,renes_approximate_2010} for related ideas). 

These ideas were 
used to construct quantum and private polar coding schemes with explicit, efficient decoders in~\cite{renes_efficient_2012} achieving rates equal to the symmetric coherent and private information, respectively, but only for a certain set of channels with essentially classical outputs (Pauli and erasure channels).
Following a different approach, Wilde and Guha~\cite{wilde_polar_2011} constructed quantum and private polar codes at these rates for any degradable channels for which the output to the environment is essentially classical. (In both cases, the private codes obey the so-called strong security criterion, such that the eavesdropper gets essentially no information about the transmitted information, not merely that she only gets information at a vanishing rate.) Both coding techniques require entanglement or secret-key assistance in the general case. 

Our new constructions have several advantages over these previous schemes:
\begin{itemize}
\item The net communication rate is equal to the symmetric coherent or private 
information for an \textit{arbitrary} quantum channel with qubit input.

\item The decoders are \textit{explicit}; in the quantum case the decoder consists of $O\left(  N\right)
$\ rounds of coherent quantum successive cancellation followed by $N$ \textsc{cnot} gates, while only an incoherent implementation of quantum successive cancellation is required in the private case. 

\item The entanglement or secret-key consumption rate vanishes for any quantum channel which is either degradable or which satisfies a certain fidelity criterion. 
\end{itemize}

Following the multi-level polar coding method of~\cite{sasoglu_polarization_2009}, we show how to extend the coding scheme to channels with multiple qubit inputs.
This gives an explicit code construction for the superactivation effect, in
which two zero-capacity channels have a
non-zero quantum capacity when used together \cite{smith_quantum_2008} (in this sense, the
channels \textit{activate} each other).

We structure this paper as follows. After setting notation and defining important quantites in Section~\ref{sec:notation}, we describe the ``amplitude'' and ``phase'' channels relevant to our coding schemes in Section~\ref{sec:complementary-channels}. In Section~\ref{sec:polar-background}, we recall the results on polarization of channels with classical input and quantum output, and in Section~\ref{sec:simultaneous}, we describe the simultaneous polarization of the amplitude and phase channels, which is the heart of our coding scheme. Sections~\ref{sec:quantumcoding} and \ref{sec:privatecoding} detail our quantum and private coding schemes, respectively. Section~\ref{sec:noassist} gives the two conditions under which the entanglement or secret-key assistance rate vanishes, while Section~\ref{sec:superactivation} outlines how to adapt our quantum
polar coding scheme so that it can exhibit the superactivation effect.
Finally, we conclude in Section~\ref{sec:conclusion}\ with a summary and some
open questions.

\section{Notation and Definitions}
\label{sec:notation}

A binary-input \textit{classical-quantum} (cq) channel $\W:x\rightarrow\rho
_{x}$ prepares a quantum state $\rho_{x}$ at the output, depending on an input
classical bit $x$. Two parameters that
determine the performance of $\W$ are the fidelity $$F\left(
\W\right)  \equiv\left\Vert \sqrt{\rho_{0}}\sqrt{\rho_{1}}\right\Vert _{1}$$
and the symmetric Holevo information $$I\left( \W\right)  \equiv H\left(
\tfrac{1}{2}\left(  \rho_{0}+\rho_{1}\right)  \right)  - \tfrac{1}{2}\left[  H\left(  \rho
_{0}\right)  +H\left(  \rho_{1}\right)  \right]  ,$$
where $\|A\|_1=\tr[\sqrt{A^\dagger A}]$ and $H\left(
\sigma\right)  \equiv-\tr\left\{  \sigma\log_{2}\sigma\right\}  $ is the von
Neumann entropy. These parameters generalize the Bhattacharya parameter and
the symmetric mutual information \cite{arikan_channel_2009}, respectively, (note that the former is denoted by $Z(\W)$ in~\cite{arikan_channel_2009}) and are related as
\cite{wilde_polar_2012-1}:
\begin{align}
I\left(  \W\right)   \approx1 &\Leftrightarrow F\left(  \W\right)  \approx0 \nonumber\\
I\left(  \W\right)  \approx0 & \Leftrightarrow F\left(  \W\right)  \approx1. \nonumber
\end{align}
The channel $\W$ is near perfect when $I\left(  \W\right)
\approx1$ and near useless when $I\left(  \W\right)  \approx0$.

A qubit-input \emph{quantum channel} $\N^{A'\rightarrow B}$ is a completely positive, trace preserving map from a two-dimensional input system $A'$ to a $d$-dimensional output system $B$. 
Every channel has an isometric extension (Stinespring dilation) to a partial isometry $U_{\N}^{A^{\prime}\rightarrow BR}$  taking $A'$ to $B$ and an additional \emph{reservoir} system $R$~\cite{wilde_classical_2011}.
Fix an arbitrary basis with elements $\left\vert z\right\rangle $ and call it the computational or ``amplitude'' basis with $z\in\left\{  0,1\right\}  $. Let $\left\vert \widetilde{x}\right\rangle $ denote the conjugate, Hadamard, or
``phase'' basis with $\widetilde{x}%
\in\left\{  +,-\right\}  $ and $\left\vert \pm\right\rangle
\equiv\left(  \left\vert 0\right\rangle \pm\left\vert 1\right\rangle \right)
/\sqrt{2}$. Furthermore, let $X$ denote the operator such that $$X\ket{z}=\ket{z\oplus 1},$$ where arithmetic inside the ket is modulo 2, and $Z$ the operator such that $$Z\ket{z}=(-1)^z\ket{z}.$$

The symmetric coherent information of a channel is defined by 
\begin{align}
\label{eq:cohinf}
I_{\rm sym}(\N)\equiv H\left(  B\right)_{\tau} -H\left(  AB\right)_{\tau},
\end{align}
where $\tau^{AB}=\N^{A'\rightarrow B}(\Phi^{AA'})$ and $\Phi^{AA'}$ denotes the maximally entangled state:
\[
\left\vert \Phi\right\rangle ^{AA^{\prime}}\equiv\tfrac{1}{\sqrt{2}}\sum
_{z\in\left\{  0,1\right\}  }\left\vert z\right\rangle ^{A}\left\vert
z\right\rangle ^{A^{\prime}}=\tfrac{1}{\sqrt{2}}\sum_{\widetilde{x}\in\left\{
+,-\right\}  }\left\vert \widetilde{x}\right\rangle ^{A}\left\vert
\widetilde{x}\right\rangle ^{A^{\prime}}.
\]

A quantum wiretap channel $\N^{A'\rightarrow BE}$ \cite{D03,1050633} is a completely positive, trace preserving map
with an input system $A^\prime$, an output system $B$ for the legitimate receiver (named Bob), and an output system
$E$ for the wiretapper (named Eve).  Appendix~\ref{app:cw-as-qw} shows that a classical wiretap channel is a special case of a quantum wiretap channel.
The symmetric private information of a quantum wiretap channel $\N^{A'\rightarrow BE}$ is defined by
\begin{align}
\label{eq:privinf}
P_{\rm sym}(\N)\equiv \max_{\rho^{ZA'}}\left[I(Z;B)_\tau-I(Z;E)_\tau\right],
\end{align}
where $\tau^{ZBE}=\N^{A'\rightarrow BE}(\rho^{ZA'})$ for $\rho^{ZA'}$ a cq state of the form: $$\rho^{ZA'}=\tfrac 12\sum_z \ket{z}\bra{z}^Z\otimes \rho_z^{A'},$$ and $\rho_z^{A'}$ an arbitrary set of (possibly mixed) states.

\section{Classical-Quantum Channels for Complementary Variables}
\label{sec:complementary-channels}

In this section we construct cq channels from a given quantum channel which will be relevant to the quantum coding procedure. Slight generalizations of these channels will be relevant to the private coding procedure. 

\subsection{CQ Channels for Quantum Communication}

Following~\cite{renes_physical_2008}, we consider building up a quantum communication
protocol from two classical communication protocols that preserve classical
information encoded into complementary variables. In this vein, two particular
classical-quantum (cq) channels are important. 
First, consider the cq channel
induced by sending an amplitude basis state over  $\N$
\begin{equation}
\W_{A}:z\rightarrow\N^{A^{\prime}\rightarrow B}\left(  \ket{z}\bra{z} \right)  \equiv\varphi^B_{z}%
,\label{eq:amplitude-cq-channel}%
\end{equation}
where the classical input $z$ is a binary variable and the notation $\W_{A}$
indicates that the classical information is encoded into the amplitude basis. We can 
regard this as the sender (Alice) modulating a standard signal $\ket{0}$ with $X^z$ and transmitting the 
result to the receiver (Bob). 

For the other cq channel, suppose that Alice instead transmits a binary variable $x$ by modulating the signal with $Z^x$, a 
rephasing of the amplitude basis states. However, instead of applying this to $\ket{0}$, she modulates one share of an entangled state $\Phi^{CA^\prime}.$ 
To transmit the binary value $x$, Alice modulates $C$ with the phase operator $Z^x$ and then sends $A'$ via the noisy channel $\N$ and $C$ via a noiseless channel to Bob. 
The overall result is the state
\begin{align}
\ket{\sigma_x}^{BCR}&=U_{\N%
}^{A^{\prime}\rightarrow BR} \left(Z^x\right)^{C}\ket{\Phi}^{A'C},\\
&=\tfrac1{\sqrt{2}}\sum_{z\in\{0,1\}}(-1)^{xz}\ket{\varphi_z}^{BR}\ket{z}^C,
\end{align} 
where $\ket{\varphi_z}^{BR}$ is a purification of $\varphi_z^{B}$ in (\ref{eq:amplitude-cq-channel}). The relevant cq channel is then of the following form:%
\begin{equation}
\W_{P}:x\rightarrow\sigma_{x}^{BC},\label{eq:phase-cq-channel}%
\end{equation}
where the notation $\W_{P}$
indicates that the classical information is encoded into a phase variable. 
In contrast to $\W_A$, the channel $\W_{P}$ is one in which the receiver has quantum side information
(in the form of system $C$) beyond what is transmitted by $\N$ itself. 
 Operationally, this quantum side information becomes
available to Bob after he coherently decodes the amplitude variable. It does
\textit{not} correspond operationally to a Bell state shared before
communication begins.

Both cq channels in (\ref{eq:amplitude-cq-channel}) and
(\ref{eq:phase-cq-channel}) arise in the error analysis of our quantum polar
coding scheme, in the sense that its performance depends on the performance of
constituent polar codes constructed for these cq channels. Moreover, the two channels
are more closely related than they may initially appear. To see their relationship, consider the ``channel state''
\begin{align}
\label{eq:channelstate}
\ket{\psi}^{ABCR}
&=\tfrac1{\sqrt{2}}{\sum_{x\in\{0,1\}}}\ket{\widetilde{x}}^A\ket{\sigma_x}^{BCR}\nonumber\\
&=\tfrac1{\sqrt{2}}{\sum_{z\in\{0,1\}}}\ket{z}^A\ket{z}^C\ket{\varphi_z}^{BR}.
\end{align} 
Measuring system $A$ in the phase basis $\ket{\widetilde{x}}$ generates the $\W_P$ output state $\sigma_x^{BC}$ in the $BC$ systems, while measuring $A$ in the amplitude basis generates the $\W_A$ output $\varphi_z^B$ in the $B$ system. 

Looking at the $R$ system output of the amplitude basis measurement defines the cq channel $\W_{R}$ to the reservoir, pertaining to amplitude information in $A$:
$$\W_{R}:z\rightarrow \varphi_z^R.$$  The uncertainty principle of~\cite{renes_conjectured_2009} then implies a relation between amplitude information about $A$ present in $R$ and phase information about $A$ present in $BC$. Indeed, due to the special form of $\ket\psi$, namely, the coherent copy of the amplitude of $A$ in system $C$, the following uncertainty relation holds~\cite{renes_physical_2008,renes_conjectured_2009}:
\begin{align}
\label{eq:certainty}
H(Z^A|R)_\psi+H(X^A|BC)_\psi = 1,
\end{align}
where $H(Z^A|B)_\psi$ is the conditional von Neumann entropy of $Z$ given $B$ for the cq state $\tfrac12\sum_z\ket{z}\bra{z}^Z\otimes\phi_z^B$ (i.e., $\psi$ after measuring $A$ in the amplitude basis), while $H(X^A|BC)_\psi$ is the conditional entropy of $X$ given $BC$ for the cq state $\tfrac12\sum_x \ket{\widetilde x}\bra{\widetilde x}^X\otimes \sigma_x^{BC}$ ($\psi$ after measuring $A$ in the phase basis). For convenience, we reproduce the proof
of \eqref{eq:certainty} in Lemma~\ref{lem:certainty} of Appendix~\ref{sec:useful}.

Since the channel inputs are presumed to be uniform, the uncertainty relation in \eqref{eq:certainty} immediately implies  
\begin{align}
\label{eq:certainty-channel}
I(\W_P)+I(\W_R)=1.
\end{align}
The more phase information goes to Bob, the less amplitude information goes to the reservoir $R$, and \emph{vice versa}. In Section~\ref{sec:ent-cons-rate-vanish} we will use this relationship to relate the 
quantum polar coding scheme presented in this article to that from prior work in~\cite{wilde_polar_2011}.

\subsection{CQ Channels for Private Communication}

For the problem of private coding in the wiretap scenario, we must also specify the eavesdropper's output, not just the intended receiver's system $B$. In general, the eavesdropper (Eve) could have access to the reservoir $R$ of $\N$ in whole \emph{or in part}. Thus, let us suppose that $R$ can be divided into two subsystems, $S$ and $E$, the latter being the output held by the eavesdropper. Clearly, $S$ does not negatively impact the security of communication between Alice and Bob. Indeed, $S$ functions as a sort of ``shield''~\cite{horodecki_secure_2005,renes_physical_2008} protecting  information in the honest parties' systems from leaking to $E$.

In the above, we have also assumed that the sender inputs a pure state to $\N$, in either the amplitude or phase basis. 
To study the private coding problem in full generality, we relax this assumption and suppose that Alice prepends an additional cq channel $\mathcal M$ to $\N$ whose job is to create (make) a state $\rho_z$ from $z$:
$$\rho_z \equiv  {\mathcal M}(\ket{z}\bra{z}) .$$ Altogether this defines the cq channel 
\begin{align}
\overline\W_A:z\rightarrow \N^{A'\rightarrow B}\circ \mathcal{M}^{A'}(\ket{z}\bra{z})\equiv\theta_z^{B}.
\end{align}
The state $\rho_z$ admits a purification $\ket{\varrho_z}^{A'S'}$ to an additional system $S'$, which functions as an additional shield system. The purification is created by a Stinespring dilation $U_{\mathcal M}^{A'\rightarrow A'S'}$ of $\mathcal M$ applied to $\ket{z}$:
\begin{align}
\ket{\varrho_z}^{A'S'}=U_{\mathcal M}^{A'\rightarrow A'S'}\ket{z}^{A'}.
\end{align}

The relevant phase channel is the same as before, with the exception that again Alice prepends $\mathcal M$ to $\N$. 
The modulation now results in 
\begin{align}
\ket{\omega_x}^{BCSS'E}&=(Z^x)^CU_{\N}^{A'\rightarrow BSE}U_{\mathcal M}^{A'\rightarrow A'S'}\ket{\Phi}^{CA'}\\
&=\tfrac{1}{\sqrt{2}}\sum_{z\in\{0,1\}}\left(  -1\right)  ^{xz}\left\vert
z\right\rangle ^{C}\left\vert \theta_{z}\right\rangle ^{BSS'E}.
\end{align}
The relevant cq channel is given by
\begin{align}
\overline{\W}_{P} : & \, x\rightarrow\omega_{x}^{BCSS'}, \label{eq:phase-channel}
\end{align}

Again the two channels $\overline{\W}_A$ and $\overline{\W}_P$ are related---the corresponding channel state (as in \eqref{eq:channelstate}) is now 
\begin{align}
\label{eq:private-channel-state}
\ket{\overline\psi}&=\tfrac1{\sqrt{2}}\sum_{x\in\{0,1\}}\ket{\widetilde{x}}^A\ket{\omega_x}^{BCSS'E}\\
&=\tfrac1{\sqrt{2}}\sum_{z\in\{0,1\}}\ket{z}^A\ket{z}^C\ket{\theta_z}^{BSS'E}.
\end{align}
The amplitude channel to the eavesdropper, $\overline{\W}_E$ is simply $\overline{\W}_E:z\rightarrow \theta_z^E$. Again the uncertainty relation in \eqref{eq:certainty} applies; it states
\begin{align}
\label{eq:certaintyprivate}
H(Z^A|E)_{\overline\psi}+H(X^A|BCSS')_{\overline\psi}=1.
\end{align}
The immediately translates into 
\begin{align}
\label{eq:private-channel-certainty}
I(\overline{\W}_P)+I(\overline{\W}_E)=1.
\end{align}
For private communication, this relation states that \textquotedblleft\textit{if the phase channel to
Bob is nearly perfect, then the amplitude channel to Eve must be nearly
useless and \emph{vice versa}}.\textquotedblright

The above uncertainty relation then enables us to construct a reliable and strongly secure polar coding scheme for 
 sending private classical data. As outlined in Section~\ref{sec:coding-scheme}, our scheme has
the sender transmit private information bits through the synthesized channels
(in the polar coding sense) that are nearly perfect in both amplitude and
phase for Bob. The fact that these synthesized amplitude channels are nearly
perfect guarantees that Bob will be able to recover these bits reliably. Meanwhile, the above uncertainty relation can be extended to the synthesized channels (see Lemma~\ref{lem:synth-unc}) and therefore the fact that the synthesized phase channels are nearly perfect for Bob guarantees
that Eve will be able to recover only a negligibly small amount of information
about the bits sent through them. 

Partitioning the synthesized channels according to amplitude and phase for Bob,
rather than according to amplitude for Bob and amplitude for Eve as in~\cite{wilde_polar_2011}, has the advantage that the scheme achieves the
symmetric private information rate for all quantum wiretap channels. Moreover,
we can prove that the secret key consumption rate vanishes for all degradable
quantum channels, and we can furthermore provide an additional sufficient condition
for when the secret key rate of the polar coding scheme vanishes.

\section{Polarization of Classical-Quantum Channels}
\label{sec:polar-background}

Wilde and Guha~\cite{wilde_polar_2012-1} demonstrated how to construct synthesized
versions of any cq channel $\W$, by channel combining and splitting \cite{arikan_channel_2009}.
For blocklength $N=2^n$, the synthesized channels are of the following
form:%
\begin{equation}
\W_{N}^{\left(  i\right)  }:u_{i}\rightarrow\rho_{\left(  i\right)  ,u_{i}%
}^{U_{1}^{i-1}B^{N}},\label{eq:split-channels}%
\end{equation}
where%
\begin{align}
\rho_{\left(  i\right)  ,u_{i}}^{U_{1}^{i-1}B^{N}} &  \equiv\sum_{u_{1}^{i-1}%
}\frac{1}{2^{i-1}}\left\vert u_{1}^{i-1}\right\rangle \left\langle u_{1}%
^{i-1}\right\vert ^{U_{1}^{i-1}}\otimes\overline{\rho}_{u_{1}^{i}}^{B^{N}},\\
\overline{\rho}_{u_{1}^{i}}^{B^{N}} &  \equiv \frac
{1}{2^{N-i}} \sum_{u_{i+1}^{N}}\rho_{u^{N}G_{N}}^{B^{N}},\\
\rho^{B^N}_{x^{N}}  & \equiv\rho^{B_1}_{x_{1}}\otimes\cdots\otimes\rho^{B_N}_{x_{N}}, 
\end{align}
and $G_{N}$ is \arikan{}'s encoding circuit matrix built from classical \textsc{cnot} gates. The interpretation of this channel is that it is the one
\textquotedblleft seen\textquotedblright\ by the input $u_{i}$ if all of the
previous bits $u_{1}^{i-1}$ are available and if we consider all the future
bits $u_{i+1}^{N}$ as randomized. This motivates the development of a quantum
successive cancellation decoder (QSCD) \cite{wilde_polar_2012-1}\ that attempts to distinguish
$u_{i}=0$ from $u_{i}=1$ by adaptively exploiting the results of previous
measurements and quantum hypothesis tests for each bit decision.

The synthesized channels $\W_{N}^{\left(  i\right)  }$ polarize, in the sense
that some become nearly perfect for classical data transmission while others
become nearly useless. To prove this result, one can model the
channel splitting and combining process as a random birth process
\cite{arikan_channel_2009,wilde_polar_2012-1}, and one can demonstrate that the induced random birth
processes corresponding to the channel parameters $I(\W_{N}^{\left(  i\right)
})$ and $F(\W_{N}^{\left(  i\right)  })$ are martingales that converge almost
surely to zero-one valued random variables in the limit of many recursions.
The following theorem from~\cite{wilde_polar_2012-1} (which uses the result in
\cite{arikan_rate_2008}) characterizes the rate at which the channel
polarization effect takes hold, and it is useful in proving
statements about the performance of polar codes for cq channels:

\begin{theorem}[Wilde \& Guha~\cite{wilde_polar_2012-1}]
\label{thm:fraction-good}For any binary input cq channel $\W$, let $\W_{2^{n}%
}^{\left(  K\right)  }$ be the random variable characterizing the $K^{\text{th}}$ split channel and $F(\W_{2^{n}}^{\left(  K\right)  })$  the fidelity of
that channel, where  $n$ indicates the level of recursion for the encoding. Then, for any
$\beta<\nicefrac12$,
\begin{align}
\lim_{n\rightarrow\infty}\Pr_{K}\{
{F(\W_{2^{n}}^{\left(  K\right)  })}<2^{-2^{n\beta}}\}=I\left(  \W\right).
\end{align}
\end{theorem}

Assuming knowledge of the good and bad channels, one can then construct a
coding scheme based on the channel polarization effect, by dividing the
synthesized channels according to the following polar coding rule:%
\begin{align}
\mathcal{G}_{N}\left(  \W,\beta\right)  & \equiv\left\{  i\in\left[  N\right]
:{F(\W_{N}^{\left(  i\right)  })}<2^{-N^{\beta}}\right\}  , \nonumber \\
\mathcal{B}_{N}\left(  \W,\beta\right)  & \equiv\left[  N\right]
\, \setminus \, \mathcal{G}_{N}\left(  \W,\beta\right)  
\label{eq:polar-coding-rule}%
\end{align}
so that $\mathcal{G}%
_{N}\left(  \W,\beta\right)  $ is the set of \textquotedblleft
good\textquotedblright\ channels and $\mathcal{B}_{N}\left(  \W,\beta\right)  $
is the set of \textquotedblleft bad\textquotedblright\ channels. The sender
then transmits the information bits through the good channels and
\textquotedblleft frozen\textquotedblright\ bits through the bad ones. A
helpful assumption for error analysis is that the frozen bits are chosen
uniformly at random such that the sender and receiver both have access to
these frozen bits. An explicit construction of a
QSCD that has an error probability scaling as
$o(2^{-\frac{1}{2}N^{\beta}})$ was provided in~\cite{wilde_polar_2012-1}. Let $\{\Lambda_{u_{\mathcal{F}^c}}^{\left(  u_{\mathcal{F}}\right)  }\}$ denote the corresponding decoding POVM, with
$u_{\mathcal{F}^c}$ the information bits and $u_{\mathcal{F}}$ the frozen bits.

The algorithm of Tal and Vardy~\cite{tal_how_2011} efficiently determines which synthesized channels are good or bad (according to a fixed fidelity or error-probability criterion), but this algorithm is not known to work for channels with quantum output. Finding an efficient code construction in the quantum case is an open problem.

\section{Simultaneous 
Polarization}
\label{sec:simultaneous}
For our quantum polar coding scheme, we utilize a coherent version of \arikan{}'s
encoder \cite{arikan_channel_2009}, meaning that the gates are quantum \textsc{cnot} gates (this is the same encoder as in~\cite{renes_efficient_2012,wilde_polar_2011}). The private polar coding scheme can simply use classical CNOT gates. Classical, amplitude-basis coding through the $N$-bit encoding circuit and
$N$ noisy channels results in an output state $\varphi_{z^NG_N}^{B^N}$ at the receiver, and the effect is to induce synthesized channels $\W_{A,N}^{\left(
i\right)  }$ as described in the previous section. Theorem~\ref{thm:fraction-good}\ states that
the fraction of amplitude-good channels (according to the criterion in
(\ref{eq:polar-coding-rule})) is equal to $I(\W_A)$ or, equivalently, $I(Z^A;B)_\psi$ using the channel state $\ket{\psi}$ from \eqref{eq:channelstate}. Again,
$Z^A$ indicates that system $A$ of $\ket\psi$ is first measured in the amplitude basis.

One of the main insights of~\cite{renes_efficient_2012} is that the same encoding operation leads to channel polarization for the phase channel $\W_P$ as well. 
In the present context, suppose Alice modulates the $C$ systems of the entangled states $\ket{\Phi}^{C^NA^{\prime N}}$ with $x^N$, but then inputs the $A^{\prime N}$ systems to the coherent encoder before
sending them via the channel to Bob. The result is  
\begin{align}
\label{eq:phaseencstate}
&
\tfrac1{\sqrt{2^N}}\!\!\!\!\sum_{z^N\in \{0,1\}^N}\!\!\!\!\!(-1)^{x^N\cdot z^N}\!\ket{\varphi_{z^NG_N}}^{B^NR^N}\!\ket{z^N}^{C^N}\\
&=
\tfrac1{\sqrt{2^N}}\!\!\!\!\sum_{z^N\in \{0,1\}^N}\!\!\!\!\!(-1)^{x^N\cdot z^NG_N}\!\ket{\varphi_{z^N}}^{B^NR^N}\!\ket{z^NG_N}^{C^N}\nonumber\\
&=
\tfrac1{\sqrt{2^N}}\!\!\!\!\sum_{z^N\in \{0,1\}^N}\!\!\!\!\!(-1)^{x^NG_N^T\cdot z^N}\!\ket{\varphi_{z^N}}^{B^NR^N}U_{\mathcal{E}}\ket{z^N}^{C^N},\nonumber
\end{align}
since $x^N\cdot z^NG_N=x^NG_N^T\cdot z^N$ and where $U_{\mathcal{E}}$ denotes the polar encoder. Thus, the $B^NC^N$ marginal state is simply $U_{\mathcal{E}}^{C^N}\sigma_{x^N G_N^T}^{B^NC^N}U_{\mathcal{E}}^{\dagger C^N}$, so that the coherent encoder also induces synthesized channels $\W_{P,N}^{(i)}$ using the encoding matrix $G_N^T$ instead of $G_N$, modulo the additional $U_{\mathcal{E}}$ acting on $C^N$. 
The fraction of phase-good channels is
approximately equal to $I(\W_P)$ or equivalently $I\left(  X^A;BC\right)_{\psi}$.

Note that the classical side information for the
$\W_{P,N}^{\left(  i\right)  }$ channels is different from that in (\ref{eq:split-channels})
 because the direction of all \textsc{cnot} gates is flipped due to the transpose of $G_N$
when acting on phase variables. The change in the direction of the \textsc{cnot}\ gates
means that the $i^{\text{th}}$ synthesized phase channel $\W_{P,N}^{\left(
i\right)  }$ is such that all of the \textit{future} bits $x_{N}\cdots
x_{i+1}$ are available to help in decoding bit $x_{i}$ while all of the
\textit{past} bits $x_{i-1}\cdots x_{1}$ are randomized. (This is the same
as described in~\cite{renes_efficient_2012}\ for Pauli channels.) 

For the case of private coding, we need only make one small modification. Instead of applying the encoding operation immediately prior to $\N$, we apply the encoding operation prior to $\N\circ \mathcal{M}$. Otherwise we proceed as before. 

\section{Quantum Coding Scheme}
\label{sec:quantumcoding}
In this section we describe the quantum coding scheme in detail. First we consider the encoder and establish the achievable rate of the protocol in the limit of infinitely-large blocklength. Then we describe the decoder and show that the protocol produces approximate ebits of fidelity exponentially close to unity between sender and receiver, justifying the rate calculation of the first subsection.
\subsection{Encoder \& Achievable Rates}

We divide the synthesized cq amplitude channels $\W_{A,N}^{\left(  i\right)  }$
into sets $\mathcal{G}_{N}\left(  \W_{A},\beta\right)  $ and $\mathcal{B}%
_{N}\left(  \W_{A},\beta\right)  $ according to (\ref{eq:polar-coding-rule}),
and similarly, we divide the synthesized cq phase channels $\W_{P,N}^{\left(
i\right)  }$ into sets $\mathcal{G}_{N}\left(  \W_{P},\beta\right)  $ and
$\mathcal{B}_{N}\left(  \W_{P},\beta\right)  $, where $\beta<\nicefrac 12$. The synthesized
channels correspond to particular inputs to the encoding operation, and thus the
set of all inputs divides into four groups:\ those 
that are good for both the amplitude and phase variable, those that are good
for amplitude and bad for phase, bad for amplitude and good for phase, and
those that are bad for both variables. We establish notation for these
channels as follows:%
\begin{align*}
\mathcal{A} &  \equiv\mathcal{G}_{N}\left(  \W_{A},\beta\right)  \,\,\cap\,\,
\mathcal{G}_{N}\left(  \W_{P},\beta\right)  ,\\
\mathcal{X} &  \equiv\mathcal{G}_{N}\left(  \W_{A},\beta\right)   \,\,\cap\,\,
\mathcal{B}_{N}\left(  \W_{P},\beta\right) \\
\mathcal{Z} &  \equiv\mathcal{B}_{N}\left(  \W_{A},\beta\right)   \,\,\cap\,\,
\mathcal{G}_{N}\left(  \W_{P},\beta\right)  ,\\
\mathcal{B} &  \equiv\mathcal{B}_{N}\left(  \W_{A},\beta\right)   \,\,\cap\,\,
\mathcal{B}_{N}\left(  \W_{P},\beta\right)  .
\end{align*}
Our quantum polar coding scheme has the sender transmit information qubits
through the inputs in $\mathcal{A}$, frozen bits in the phase basis through
the inputs in $\mathcal{X}$, frozen bits in the amplitude basis through the
inputs in $\mathcal{Z}$, and halves of ebits shared with Bob through the inputs in
$\mathcal{B}$ (we can think of these in some sense as being frozen
simultaneously in both the amplitude and phase basis). 

Thus, our coding procedure is \emph{entanglement-assisted} \cite{brun_correcting_2006}.
Indeed, the encoder implicitly results in an 
entanglement-assisted Calderbank-Shor-Steane (CSS) code, as pointed out in~\cite{renes_efficient_2012}.
In the stabilizer language of quantum error-correcting codes, the values
of the frozen inputs determine the various stabilizers of the code, and due
to the dual nature of the encoding circuit (polarizing both amplitude and phase
inputs), frozen amplitude states become $Z$-type stabilizers and frozen phase states become $X$-type stabilizers. The need for entangled inputs signals that the CSS code is entanglement-assisted. As some inputs require both frozen amplitude and phase values, the resulting stabilizer code would need both the corresponding $X$- and $Z$-type stabilizers. These, however, do not commute, and the role of entanglement-assistance is to ``enlarge'' the stabilizers to additional systems on Bob's side such that they do commute.
In spite of the fact that our quantum polar coding scheme results in a CSS code, the decoding procedure
(quantum successive cancellation decoding) is very different from the standard stabilizer recovery procedure in which the receiver performs stabilizer measurements
and classical post processing of syndromes.

The net rate of the protocol for blocklength $N$ is simply $r_Q(N) \equiv \frac{|\mathcal A|-|\mathcal B|}{N}$. Upon suitable choice of amplitude basis, it equals the symmetric coherent information in the asymptotic limit:
\begin{theorem}
\label{thm:qrate}
$\lim_{N\rightarrow \infty}r_Q(N)=I_{\rm sym}(\N).$
\end{theorem}
\begin{IEEEproof}
From Theorem~\ref{thm:fraction-good} it follows that 
\begin{align}&\lim_{N\rightarrow\infty
}\frac{1}{N}\left\vert \mathcal{G}_{N}\left( \W_A,\beta\right)  \right\vert
=I\left(  Z^A;B\right)_\psi,\qquad\text{and}\\&\lim_{N\rightarrow\infty}\frac{1}{N}\left\vert
\mathcal{G}_{N}\left(\W_P,\beta\right)  \right\vert =I\left(  X^A;BC\right)_\psi.
\end{align}
From basic set theory we have
\begin{align*}
\left\vert \mathcal{A}\right\vert   
  =&\left\vert \mathcal{G}_{N}\left(  \W_{A},\beta\right)  \right\vert
+\left\vert \mathcal{G}_{N}\left(  \W_{P},\beta\right)  \right\vert\\
& -\left\vert
\mathcal{G}_{N}\left(  \W_{A},\beta\right)  \,\,\cup\,\,\mathcal{G}_{N}\left(
\W_{P},\beta\right)  \right\vert,
\end{align*}
as well as
\begin{align*}
\left\vert \mathcal{G}_{N}\left( \W_A,\beta\right)  \,\,\cup\,\,\mathcal{G}%
_{N}\left( \W_P,\beta\right)  \right\vert  = N-\left\vert \mathcal{B}\right\vert .
\end{align*}
Thus, the rate of the scheme is equal to 
\begin{align}
\lim_{N\rightarrow\infty}\frac{|\mathcal A|-|\mathcal B|}{N}  &
=  I\left(  Z^A;B\right)_\psi  +I\left(  X^A;BC\right)_\psi  -1 
\label{eq:net-rate-calc}\\
&  =  I\left(  Z^A;B\right)_\psi  -I\left(  Z^A;R\right)_\psi \nonumber\\
&  = H\left(  B\right)_\psi - H\left( B| Z^A\right)_\psi  \nonumber \\
& \,\,\,\,\,\,\,\, - [H\left(  R\right)_\psi - H\left( R| Z^A\right)_\psi] \nonumber\\
&  = H\left(  B\right)_\psi  - H\left(  R\right)_\psi  \nonumber\\
&  =H(B)_\tau-H(AB)_\tau.\nonumber
\end{align}
Here the second equality uses \eqref{eq:certainty-channel}.
The third equality is an identity, and the fourth follows from the fact
that the state on $BR$ is pure when conditioned on a measurement outcome of $A$,
so that $H\left( B| Z^A\right)_\psi = H\left( R| Z^A\right)_\psi$.
The final equality uses the fact that the entropy expressions
are equal when evaluated for the state $\tau^{AB}=\N^{A'\rightarrow B}(\Phi^{AA'})$. 
\end{IEEEproof}

\subsection{Decoder and Error Analysis}

\label{sec:err-analysis}
We now describe the decoder in more detail and demonstrate that the fidelity of the entire coding scheme
becomes exponentially close to one as the blocklength gets large. In particular,  we will show
\begin{theorem}
\label{thm:entquality}
Given any quantum channel with a qubit input system and a finite-dimensional output system, for large enough blocksize $N$, there exists a quantum polar coding scheme which generates approximate ebits
that are $o(2^{-\frac 12N^\beta})$-close in trace distance to exact ebits  for $\beta<\nicefrac12$. Furthemore,
the rate of this scheme is equal to the symmetric coherent information of the channel.
\end{theorem}
\begin{IEEEproof}
The sender and receiver begin with the following state:%
\begin{align*}
\ket{\Psi_0}=
N_{0}\sum_{u_{\mathcal{A}},u_{\mathcal{B}}}\left\vert u_{\mathcal{A}%
}\right\rangle \left\vert u_{\mathcal{A}}\right\rangle \left\vert
u_{\mathcal{Z}}\right\rangle \left\vert \widetilde{u}_{\mathcal{X}}\right\rangle
\left\vert u_{\mathcal{B}}\right\rangle\otimes \left\vert u_{\mathcal{B}%
}\right\rangle ,
\end{align*}
where Alice possesses the first five registers, Bob
the last one,\footnote{In quantum information theory
 the tensor product symbol is often used implicitly. Our convention in this section is to leave
it implicit for systems belonging to the same party and use it explicitly 
to denote a division between two parties.} and $N_{0}\equiv1/\sqrt{2^{\left\vert \mathcal{A}\right\vert
+\left\vert \mathcal{B}\right\vert }}$. We also assume for now that the bits
in $u_{\mathcal{Z}}$ and $u_{\mathcal{X}}$ are chosen uniformly at random and
are known to both the sender and receiver. Note that the fourth register is expressed in the phase basis; using the amplitude 
basis instead gives 
\begin{align*}
\ket{\Psi_0}=
N_{1}\hspace{-3mm}\sum_{u_{\mathcal{A}},u_{\mathcal{B}},v_{\mathcal{X}}}\hspace{-3mm}\left(  -1\right)
^{u_{\mathcal{X}}\cdot v_{\mathcal{X}}}\!\left\vert u_{\mathcal{A}}\right\rangle
\left\vert u_{\mathcal{A}}\right\rangle \left\vert u_{\mathcal{Z}%
}\right\rangle \left\vert v_{\mathcal{X}}\right\rangle \left\vert
u_{\mathcal{B}}\right\rangle \otimes \left\vert u_{\mathcal{B}}\right\rangle ,
\end{align*}
where 
 $N_{1}\equiv
1/\sqrt{2^{\left\vert \mathcal{A}\right\vert +\left\vert \mathcal{B}%
\right\vert +\left\vert \mathcal{X}\right\vert }}$. The sender then feeds the
middle four registers through the polar encoder and channel, leading to a state of the following form:%
\begin{align*}
\ket{\Psi_1}=
N_{1}\hspace{-3mm}\sum_{u_{\mathcal{A}},u_{\mathcal{B}},v_{\mathcal{X}}}\hspace{-3mm}\left(  -1\right)
^{u_{\mathcal{X}}\cdot v_{\mathcal{X}}}\!\left\vert u_{\mathcal{A}}\right\rangle\otimes
\left\vert \varphi_{u_{\mathcal{A}},u_{\mathcal{Z}},v_{\mathcal{X}}%
,u_{\mathcal{B}}}\right\rangle ^{B^NR^N}\!\left\vert u_{\mathcal{B}%
}\right\rangle ,
\end{align*}
where $\left\vert \varphi_{u_{\mathcal{A}},u_{\mathcal{Z}},v_{\mathcal{X}%
},u_{\mathcal{B}}}\right\rangle ^{B^NR^N}\equiv U_{\mathcal{N}%
}^{\otimes N}U_{\mathcal{E}%
}\left\vert u_{\mathcal{A}}\right\rangle \left\vert u_{\mathcal{Z}%
}\right\rangle \left\vert v_{\mathcal{X}}\right\rangle \left\vert
u_{\mathcal{B}}\right\rangle $ (abusing notation, the encoding operation $G_N$ is left implicit).

Observe that, conditioned on amplitude measurements of $\ket{u_{\mathcal{A}}}$ and $\ket{u_{\mathcal{B}}}$, the $B^N$ subsystem is identical to the polar-encoded
output of $\W_A$. Thus, the first step of
the decoder is the following coherent implementation of the QSCD for the amplitude channel $\W_A$ 
\begin{align}
V_A=\!\!\!\!\!\sum_{u_{\mathcal{A}},u_{\mathcal{B}},v_{\mathcal{X}}}\!\!\!\sqrt{\Lambda
_{u_{\mathcal{A}},v_{\mathcal{X}}}^{\left(  u_{\mathcal{B}},u_{\mathcal{Z}%
}\right)  }}\otimes\left\vert u_{\mathcal{A}}\right\rangle \left\vert
v_{\mathcal{X}}\right\rangle \otimes\left\vert u_{\mathcal{B}}\right\rangle
\left\vert u_{\mathcal{B}}\right\rangle \left\langle u_{\mathcal{B}%
}\right\vert \otimes\left\vert u_{\mathcal{Z}}\right\rangle
.\label{eq:amp-decoder}%
\end{align}
The idea here is that the decoder coherently recovers the bits in
$u_{\mathcal{A}}$ and $v_{\mathcal{X}}$, using $u_{\mathcal{Z}}$
and $u_{\mathcal{B}}$ as classical and quantum side information, respectively.
Appendix~\ref{app:err-analysis-amplitude} provides a detailed argument that the state resulting from this first decoding step
is $o(2^{-\frac 12N^{\beta}})$-close in expected
trace distance to the following ideal state:
\begin{align}
\ket{\Psi_2}=N_{1}\hspace{-2mm}\sum_{u_{\mathcal{A}},u_{\mathcal{B}},v_{\mathcal{X}}}\hspace{-2mm}\left(  -1\right)
^{u_{\mathcal{X}}\cdot v_{\mathcal{X}}}  & \left\vert u_{\mathcal{A}}\right\rangle
\left\vert \varphi_{u_{\mathcal{A}},u_{\mathcal{Z}},v_{\mathcal{X}}%
,u_{\mathcal{B}}}\right\rangle ^{B^{N}R^{N}}\otimes\nonumber\\
&\left\vert u_{\mathcal{A}}\right\rangle \left\vert v_{\mathcal{X}%
}\right\rangle \left\vert u_{\mathcal{B}}\right\rangle \left\vert
u_{\mathcal{B}}\right\rangle \left\vert u_{\mathcal{Z}}\right\rangle ,
\label{eq:ampidealstate}
\end{align}
where the expectation is with respect to the uniformly random choice of 
$u_{\mathcal{X}}$. Thus, Bob has coherently recovered the bits
$u_{\mathcal{A}}$ and $v_{\mathcal{X}}$ with the decoder in
(\ref{eq:amp-decoder}), while making a second coherent and incoherent copy of
the bits $u_{\mathcal{B}}$ and $u_{\mathcal{Z}}$, respectively.

The next step in the process is to make coherent use of the $\W_P$ decoder. For this to be useful, however, 
we must show that encoded versions of $\ket{\sigma_x}^{BCE}$, as in (\ref{eq:phaseencstate}), are present in $\ket{\Psi_2}$. To see this, first observe that we can write 
\begin{multline*}
\ket{\Psi_2}=
N_{2}\sum_{\substack{u_{\mathcal{A}},u_{\mathcal{B}},v_{\mathcal{X}%
},\\{x}_{\mathcal{A}},{x}_{\mathcal{B}}}}\left(
-1\right)  ^{u_{\mathcal{X}}\cdot v_{\mathcal{X}}+{x}_{\mathcal{A}%
}\cdot u_{\mathcal{A}}+{x}_{\mathcal{B}}\cdot u_{\mathcal{B}}%
}\left\vert \widetilde{x}_{\mathcal{A}}\right\rangle \otimes\\
\left\vert \varphi_{u_{\mathcal{A}},u_{\mathcal{Z}},v_{\mathcal{X}}%
,u_{\mathcal{B}}}\right\rangle ^{B^{N}R^{N}}\left\vert u_{\mathcal{A}%
}\right\rangle \left\vert v_{\mathcal{X}}\right\rangle \left\vert
u_{\mathcal{B}}\right\rangle \left\vert \widetilde{x}_{\mathcal{B}%
}\right\rangle \left\vert u_{\mathcal{Z}}\right\rangle ,
\end{multline*}
where $N_{2}\equiv1/\sqrt{2^{2\left\vert \mathcal{A}\right\vert +2\left\vert
\mathcal{B}\right\vert +\left\vert \mathcal{X}\right\vert }}$, by expressing the first register and the second $\left\vert u_{\mathcal{B}%
}\right\rangle $ register in the phase basis. 
This is nearly the expression we are looking for, as all the desired phase factors are present, except one corresponding to  $\ket{u_{\mathcal{Z}}}$. 

As $u_{\mathcal{Z}}$ is chosen at random, we can describe it quantum-mechanically as arising from part of an entangled state. The other part is shared by Alice and an inaccessible reference system. Including this purification degree of freedom, $\ket{\Psi_2}$ becomes
\begin{align*}
\ket{\Psi_2'}=N_{3}\sum_{\substack{u_{\mathcal{A}},u_{\mathcal{B}},v_{\mathcal{X}%
},\\u_{\mathcal{Z}},{x}_{\mathcal{A}},{x}_{\mathcal{B}}}}\left(
-1\right)  ^{u_{\mathcal{X}}\cdot v_{\mathcal{X}}+{x}_{\mathcal{A}%
}\cdot u_{\mathcal{A}}+{x}_{\mathcal{B}}\cdot u_{\mathcal{B}}%
}\left\vert \widetilde{x}_{\mathcal{A}}\right\rangle \otimes\\
\left\vert \varphi_{u_{\mathcal{A}},u_{\mathcal{Z}},v_{\mathcal{X}}%
,u_{\mathcal{B}}}\right\rangle ^{B^{N}E^{N}}\left\vert u_{\mathcal{A}%
}\right\rangle \left\vert v_{\mathcal{X}}\right\rangle \left\vert
u_{\mathcal{B}}\right\rangle \left\vert \widetilde{x}_{\mathcal{B}%
}\right\rangle \left\vert u_{\mathcal{Z}}\right\rangle\otimes \ket{u_{\mathcal{Z}}} ,
\end{align*}
where $N_3=N_2/\sqrt{2^{|\mathcal{Z}|}}$. Again utilizing the phase basis gives 
\begin{align*}
\ket{\Psi_2'}=N_{3}\hspace{-3mm}\sum_{\substack{u_{\mathcal{A}},u_{\mathcal{B}},v_{\mathcal{X}%
},u_{\mathcal{Z}},\\{x}_{\mathcal{A}},{x}_{\mathcal{B}},{x}_{\mathcal{Z}}}}\hspace{-3mm}\left(
-1\right)  ^{u_{\mathcal{X}}\cdot v_{\mathcal{X}}+{x}_{\mathcal{A}%
}\cdot u_{\mathcal{A}}+{x}_{\mathcal{B}}\cdot u_{\mathcal{B}}+{x}_{\mathcal{Z}}\cdot u_{\mathcal{Z}}%
}\left\vert \widetilde{x}_{\mathcal{A}}\right\rangle \otimes\\
\left\vert \varphi_{u_{\mathcal{A}},u_{\mathcal{Z}},v_{\mathcal{X}}%
,u_{\mathcal{B}}}\right\rangle ^{B^{N}E^{N}}\left\vert u_{\mathcal{A}%
}\right\rangle \left\vert v_{\mathcal{X}}\right\rangle \left\vert
u_{\mathcal{B}}\right\rangle \left\vert \widetilde{x}_{\mathcal{B}%
}\right\rangle \left\vert u_{\mathcal{Z}}\right\rangle\otimes\ket{\widetilde{x}_{\mathcal{Z}}}.
\end{align*}
Thus, $\ket{\Psi_2'}$ is a superposition of polar encoded states as in (\ref{eq:phaseencstate}) and therefore the phase decoder will be useful to the receiver. In particular, Bob can first apply $U_{\mathcal{E}}^{\dagger C^N}$ 
and then coherently apply the QSCD for the phase channel $\W_P$,
\begin{align}
V_P=
\!\!\!\sum_{{x}_{\mathcal{A}},x_{\mathcal{Z}},{x}_{\mathcal{B}}%
}\!\!\!\sqrt{\Gamma_{{x}_{\mathcal{A}},x_{\mathcal{Z}}}^{\left(
{x}_{\mathcal{B}},u_{\mathcal{X}}\right)  }}\otimes\left\vert
\widetilde{x}_{\mathcal{A}}\right\rangle \left\vert \widetilde{x}_{\mathcal{Z}%
}\right\rangle \left\vert \widetilde{u}_{\mathcal{X}}\right\rangle \otimes\left\vert
\widetilde{x}_{\mathcal{B}}\right\rangle \left\langle \widetilde
{x}_{\mathcal{B}}\right\vert
\label{eq:phasedecoder}
\end{align}
to coherently extract the values of ${x}_{\mathcal{A}}$ and $x_{\mathcal{Z}}$ using the frozen bits ${x}_{\mathcal{B}}$ and  ${u}_{\mathcal{X}}$. He then applies $U_{\mathcal{E}}^{C^N}$ to restore the $C^N$ registers to their previous form.  As with the amplitude decoding step, a similar argument (detailed in Appendix~\ref{app:err-analysis-phase})
ensures the closeness of the output of this process
to the ideal output as governed by the error probability of the $\W_P$ decoder. 
To express the ideal output succinctly, we first make the assignments  
\begin{align*}
\left\vert \Phi_{\mathcal{A}}\right\rangle  &  \equiv\frac{1}{\sqrt
{2^{\left\vert \mathcal{A}\right\vert }}}\sum_{u_{\mathcal{A}}}\left\vert
u_{\mathcal{A}}\right\rangle \left\vert u_{\mathcal{A}}\right\rangle
,\ \left\vert \Phi_{\mathcal{Z}}\right\rangle \equiv\frac{1}{\sqrt
{2^{\left\vert \mathcal{Z}\right\vert }}}\sum_{v_{\mathcal{Z}}}\left\vert
v_{\mathcal{Z}}\right\rangle \left\vert v_{\mathcal{Z}}\right\rangle ,\\
\left\vert \Phi_{\mathcal{X}}\right\rangle  &  \equiv\frac{1}{\sqrt
{2^{\left\vert \mathcal{X}\right\vert }}}\sum_{v_{\mathcal{X}}}\left\vert
v_{\mathcal{X}}\right\rangle \left\vert v_{\mathcal{X}}\right\rangle
,\ \left\vert \Phi_{\mathcal{B}}\right\rangle \equiv\frac{1}{\sqrt
{2^{\left\vert \mathcal{B}\right\vert }}}\sum_{u_{\mathcal{B}}}\left\vert
u_{\mathcal{B}}\right\rangle \left\vert u_{\mathcal{B}}\right\rangle.
\end{align*}
Rewriting phase terms with Pauli operators, we then have that the actual output of this step of the decoder is $o(2^{-\frac 12 N^{\beta}})$-close in expected trace distance to the following ideal state:
\begin{align}
&\ket{\Psi_3}=N_4\hspace{-3mm}\sum_{x_{\mathcal{A}},x_{\mathcal{B}},x_{\mathcal{Z}}}\hspace{-3mm}\ket{\widetilde{x}_{\mathcal{A}}}\otimes 
\ket{\widetilde{x}_{\mathcal{A}}}\ket{\widetilde{x}_{\mathcal{Z}}}
\ket{\widetilde{u}_{\mathcal{X}}}\ket{\widetilde{x}_{\mathcal{B}}}\nonumber\\
&\quad Z^{x_{\mathcal{A}},x_{\mathcal{Z}},u_{\mathcal{X}},x_{\mathcal{B}}}U_{\mathcal{N}}^{\otimes N}U_{\mathcal{E}}\ket{\Phi_{\mathcal{A}}} \ket{\Phi_{\mathcal{Z}}}\ket{\Phi_{\mathcal{X}}}\ket{\Phi_{\mathcal{B}}}\otimes \ket{\widetilde{x}_{\mathcal{Z}}},
\label{eq:ideal-phase-state}
\end{align}
where $N_4 \equiv 1/\sqrt{2^{|\mathcal{A}|+|\mathcal{B}|+|\mathcal{Z}|}}$.
Here $Z^{x_{\mathcal{A}},x_{\mathcal{Z}},u_{\mathcal{X}},x_{\mathcal{B}}}$ is shorthand for $Z^{x_{\mathcal{A}}}\otimes Z^{x_{\mathcal{Z}}}\otimes Z^{u_{\mathcal{X}}}\otimes Z^{x_{\mathcal{B}}}$, which acts on the second qubits in the entangled pairs, while the encoding and channel unitaries act on the first. 

The final step in the decoding process is to remove (or ``decouple'') the phase operator $Z^{x_{\mathcal{A}},x_{\mathcal{Z}},v_{\mathcal{X}},x_{\mathcal{B}}}$ by controlled operations from the registers $\ket{\widetilde{x}_{\mathcal{A}}}\ket{\widetilde{x}_{\mathcal{Z}}}\ket{\widetilde{u}_{\mathcal{X}}}\ket{\widetilde{x}_{\mathcal{B}}}$ to the second qubits in the entangled pairs. This phase-basis controlled phase operation is equivalent to $N$ \textsc{cnot} operations from the latter systems to the former and results in
\begin{align*}%
\!\!\!\!N_0\sum_{x_{\mathcal{A}}}\ket{\widetilde{x}_{\mathcal{A}}}\otimes \ket{\widetilde{x}_{\mathcal{A}}} U_{\mathcal{N}}^{\otimes N}U_{\mathcal{E}}\ket{\Phi_{\mathcal{A},\mathcal{Z},\mathcal{X},\mathcal{B}}}
\sum_{x_{\mathcal{B}}}\ket{u_{\mathcal{X}}}\ket{\widetilde{x}_{\mathcal{B}}},
\end{align*}
with Bob sharing $1/\sqrt{2^{|\mathcal{Z}|}}\sum_{x_{\mathcal{Z}}} \ket{\widetilde{x}_{\mathcal{Z}}}\otimes \ket{\widetilde{x}_{\mathcal{Z}}}$ with the inaccessible reference.
Thus, applying the triangle inequality for trace distance, the protocol finishes with Alice and Bob
sharing a state close in trace distance to the following state:
$$ \frac{1}{\sqrt{|\mathcal{A}|}}\sum_{x_{\mathcal{A}}}\ket{\widetilde{x}_{\mathcal{A}}}\otimes \ket{\widetilde{x}_{\mathcal{A}}}, $$
That is, the sender and receiver generate $|\mathcal{A}|$ approximate ebits with trace distance less than $o(2^{-\frac12N^{\beta}})$ to ideal ebits at the end of the protocol.
\end{IEEEproof}

\begin{remark}
The above scheme performs well with respect to a uniformly random choice of
the bits $u_{\mathcal{X}}$ and $u_{\mathcal{Z}}$, in the sense that the
expectation of the fidelity is high. However, Markov's inequality implies that a large fraction of the possible codes have good performance.
\end{remark}

\begin{remark}
The first step of the decoder is identical to the first step of the decoder
from~\cite{wilde_polar_2011}. Though, the second step above is an improvement over
the second step in~\cite{wilde_polar_2011} because it is an explicit coherent
QSCD, rather than an inexplicit
controlled-decoupling unitary.
Additionally, the decoder's complexity is equivalent to $O\left(
N\right)  $\ quantum hypothesis tests and other unitaries resulting from the
 polar decompositions of
$\Lambda_{u_{\mathcal{A}},v_{\mathcal{X}}}^{\left(  u_{\mathcal{B}},u_{\mathcal{Z}}\right)}$ and
$\Gamma_{{x}_{\mathcal{A}},x_{\mathcal{Z}}}^{\left({x}_{\mathcal{B}},u_{\mathcal{X}}\right)}$,
but it remains unclear how to
implement these efficiently.
\end{remark}

\section{Private Coding Scheme}
\label{sec:privatecoding}
The private coding scheme is a slight variation of the quantum coding scheme and makes use of the fact that the quantum encoder is based on a CSS code. Indeed, one can reduce the quantum case to a classical protocol, as was first demonstrated by Shor and Preskill~\cite{shor_simple_2000}. Renes {\it et al}.~\cite{renes_efficient_2012} point out that this reduction  applies to quantum polar codes, being CSS codes. Here we shall follow a more direct route by defining and analyzing the private coding procedure independently of the quantum protocol. 
\subsection{Encoder \& Achievable Rates}
\label{sec:coding-scheme}
As before, we divide the encoder inputs  into four groups 
\begin{align*}
\mathcal{A}  &  \equiv\mathcal{G}_{N}\left(  \overline{\W}_{A},\beta\right)
\,\,\cap\,\,\mathcal{G}_{N}\left(  \overline{\W}_{P},\beta\right)  ,\\
\mathcal{X}  &  \equiv\mathcal{G}_{N}\left(  \overline{\W}_{A},\beta\right)
\,\,\cap\,\,\mathcal{B}_{N}\left(  \overline{\W}_{P},\beta\right)  ,\\
\mathcal{Z}  &  \equiv\mathcal{B}_{N}\left(  \overline{\W}_{A},\beta\right)
\,\,\cap\,\,\mathcal{G}_{N}\left(  \overline{\W}_{P},\beta\right)  ,\\
\mathcal{B}  &  \equiv\mathcal{B}_{N}\left(  \overline{\W}_{A},\beta\right)
\,\,\cap\,\,\mathcal{B}_{N}\left(  \overline{\W}_{P},\beta\right).
\end{align*}
Unlike the quantum coding scheme, all inputs are now made in the amplitude basis. 
The sender again inputs information bits to $\mathcal A$ and frozen bits into $\mathcal Z$.
Now the sender mimics frozen phase inputs to $\mathcal X$ with random bits and mimics halves of ebits input to $\mathcal B$ with secret key bits. Thus, our private coding procedure is generally \emph{secret-key assisted}. Its rate is simply $r_P(N)=\frac{|\mathcal A|-|\mathcal B|}{N}$. It equals the symmetric private information in the asymptotic limit:
\begin{theorem}
$\lim_{N\rightarrow \infty}r_P(N)=P_{\rm sym}(N).$
\end{theorem}
\begin{IEEEproof}
The proof is entirely similar to the proof of Theorem~\ref{thm:qrate}, with the exception that we choose $\mathcal{M}$ to create the optimal states $\rho_z$ in the symmetric private information. Then we use 
\begin{align}
&\lim_{N\rightarrow\infty
}\frac{1}{N}\left\vert \mathcal{G}_{N}\left( \overline{\W}_A,\beta\right)  \right\vert
=I\left(  Z^A;B\right)_{\overline\psi},\qquad\\&\lim_{N\rightarrow\infty}\frac{1}{N}\left\vert
\mathcal{G}_{N}\left(\overline{\W}_P,\beta\right)  \right\vert =I\left(  X^A;BCSS'\right)_{\overline\psi},
\end{align}
and appeal to \eqref{eq:private-channel-certainty} instead of \eqref{eq:certainty-channel}.
\end{IEEEproof}

We should stress that our consideration of the phase channels in this part of the paper is
only necessary in order to compute the index sets $\mathcal{A}$, $\mathcal{X}%
$, $\mathcal{Z}$, and $\mathcal{B}$. The decoder in the next section does not
make explicit use of these phase channels---they only arise in our security
analysis, where we appeal to the uncertainty relation in \eqref{eq:certaintyprivate} in order to
guarantee security of the scheme. This is in contrast to the quantum polar
coding scheme of the previous sections, in which the
decoder makes explicit use of the phase channels.

\subsection{Decoder \& Reliability and Security Analysis}

We now describe the decoder and examine the reliability and security of the resulting protocol. Our approach leads to strong security, as stated in the following theorem:
\begin{theorem}
For sufficiently large $N$, the private polar coding scheme given above satisfies the following reliability and security criteria:
\begin{enumerate}
\item $\Pr\{\widehat{U}_{\mathcal{C}}\neq U_{\mathcal{C}}\}\leq o\left(
2^{-\frac{1}{2}N^{\beta}}\right)$, \, for $\mathcal{C}\equiv \mathcal{A}\cup\mathcal X$, and
\item $I(U_{\mathcal{A}};E^N)\leq  o\left(
2^{-\frac{1}{2}N^{\beta}}\right)$.
\end{enumerate}
\end{theorem}
\begin{IEEEproof}
First, it is straightforward to prove that the code has good reliability, by
appealing to the results from~\cite{wilde_polar_2012-1}. That is, there exists a
POVM $\{\Lambda_{u_{\mathcal{A}},u_{\mathcal{X}}}^{\left(  u_{\mathcal{Z}%
}, u_{\mathcal{B}%
}\right)  }\}$, the quantum successive cancellation decoder, such that%
\begin{align}
\Pr\{\widehat{U}_{\mathcal{C}}\neq U_{\mathcal{C}}\}& \leq\sqrt{2\sum
_{i\in\mathcal{C}}{F(\W_{A,N}^{\left(  i\right)  })}}\\
& =o\left(
2^{-\frac{1}{2}N^{\beta}}\right)  .
\end{align}
where $\mathcal{C\equiv A}\cup\mathcal{X}$.  The QSCD operates exactly as in~\cite{wilde_polar_2012-1}, treating $\mathcal Z\cup \mathcal B$ as the frozen set and decoding all bits in the set $\mathcal A\cup \mathcal X$, though of course only the bits in $\mathcal A$ contain the transmitted message. 
This decoder has an
efficient implementation if the channel to Bob is classical \cite{arikan_channel_2009}. This
is the case for the amplitude damping channel, the erasure channel, and any Pauli channel, for example.

We now prove that strong security, in the sense of~\cite{mahdavifar_achieving_2011}, holds for
our polar coding scheme.
Consider that%
\begin{align*}
I\left(  U_{\mathcal{A}};E^{N}\right)   &  =\sum_{i\in\mathcal{A}}I\left(
U_{i};E^{N}\big|U_{\mathcal{A}_{i}^{-}}\right)  \\
& =\sum_{i\in\mathcal{A}}I\left(
U_{i};E^{N}U_{\mathcal{A}_{i}^{-}}\right) \\
&  \leq\sum_{i\in\mathcal{A}}I\left(  U_{i};E^{N}U_{1}^{i-1}\right)\\
& =\sum_{i\in\mathcal{A}}I(\overline{\W}_{E,N}^{\left(  i\right)  })
\end{align*}
The first equality is from the chain rule for quantum mutual information and
by defining $\mathcal{A}_{i}^{-}$ to be the indices in $\mathcal{A}$ preceding
$i$. The second equality follows from the assumption that the bits in
$U_{\mathcal{A}_{i}^{-}}$ are chosen uniformly at random. The first inequality
is from quantum data processing. The third equality is from the definition of
the synthesized channels $\overline{\W}_{E,N}^{\left(  i\right)  }$. Continuing, we have%
\begin{align*}
&  \leq\sum_{i\in\mathcal{A}}\sqrt{1-F(\overline{\W}_{E,N}^{\left(  i\right)  })^2} \\
& \leq\sum_{i\in\mathcal{A}}\sqrt{1-(1-2F(\overline{\W}_{P,N}^{\left(  i\right)  }%
))^{2}}\\
&  \leq\sum_{i\in\mathcal{A}}\sqrt{4{F(\overline{\W}_{P,N}^{\left(  i\right)  })}%
}\\
&\leq2\sum_{i\in\mathcal{A}}\sqrt{2^{-N^{\beta}}} \\
& =o\left(  2^{-\frac{1}%
{2}N^{\beta}}\right)  .
\end{align*}
The first inequality is from~\cite[Proposition~1]{wilde_polar_2012-1}. The second holds since the two synthesized channels obey an uncertainty relation, shown in Lemma~\ref{lem:synth-unc}, which then gives a fidelity uncertainty relation, shown in Lemma~\ref{lem:channelfidelityunc}. Here we set $\W_1=\overline{\W}_{P,N}^{(i)}$ and $\W_2=\overline{\W}_{E,N}^{(i)}$ in the latter lemma. The fourth inequality follows from
the definition of the set $\mathcal{A}$.
\end{IEEEproof}

A clear advantage of the current approach over the
previous construction from~\cite{wilde_polar_2011}\ is that
Theorem~\ref{thm:fraction-good} directly applies to the phase-good channels
with the \textquotedblleft goodness criterion\textquotedblright\ given by
(\ref{eq:polar-coding-rule}). Only the amplitude
channels to Eve (rather than the phase-good channels to Bob) were considered in~\cite{wilde_polar_2011}, and it seemed
only possible to prove polarization results for quantum wiretap channels in
which the amplitude channel to Eve is classical. Our approach here overcomes
this difficulty by appealing to Theorem~\ref{thm:fraction-good} directly for
polarization and later relating the phase channels to Bob and the amplitude
channels to Eve via an uncertainty relation.

\section{Entanglement \& Secret-Key Assistance Are Not Always Needed}
\label{sec:noassist}
In this section we give two different conditions for when the quantum or private coding schemes require only a sublinear amount of entanglement or secret-key assistance. The first occurs when the channel is degradable~\cite{devetak_capacity_2005}, in the sense that the $R$ output (Eve's output $E$ in the private case) can be generated from Bob's output $B$ by some quantum operation. The second stems from properties of the binary erasure channel (BEC) as an ``extreme point'' in the channel synthesis process. 
\subsection{Degradable Channels}
\label{sec:ent-cons-rate-vanish}
Suppose that the channel $\mathcal{N}$ is degradable in the sense that the $R$ output $\varphi_z^{R}$ can be generated from the $B$ output $\varphi_z^B$ by some other quantum channel $\mathcal{D}$: $\varphi_z^{R}=\mathcal{D}(\varphi_z^B)$. Similarly, in the private coding scenario, suppose that $\varrho_z^E=\mathcal{D}(\varrho_z^B)$. Then we can show
\begin{theorem}
\label{thm:degradable}
For degradable channels in the quantum or private coding scenearios as described above,
the rate of entanglement or secret key assistance, respectively, vanishes in the limit of large blocklength:
\begin{align}
\lim_{N\rightarrow\infty}\tfrac1N|\mathcal{B}|=0.
\end{align}
\end{theorem}

We provide a brief summary of the proof and then follow with more detail. 
First, the channel uncertainty relation~\eqref{eq:certainty-channel} can be extended to synthesized channels
(as the inequality given in Lemma~\ref{lem:synth-unc}) and implies that phase-good channels (output fidelity near zero) to Bob are amplitude-``very bad'' channels to $R$ (output fidelity near one). From
degradability, we also know that the doubly-bad channels 
$\mathcal{B}$ are amplitude-bad channels to $R$. These two observations imply
that the doubly-bad channels to Bob, the phase-good channels to Bob, and
amplitude-good channels to $R$ are disjoint sets. The sum of the fractional sizes of the latter two sets equals $I\left(
X^A;BC\right)_\psi  +I\left(  Z^A;R\right)_\psi  =1$ by~\eqref{eq:certainty}, implying that the rate of the
doubly-bad channel set $\mathcal{B}$ (the entanglement consumption rate) approaches zero in
the same limit. Replaying the argument for the channels in the private communication scenario gives the same result.

\begin{IEEEproof}
The proof is a
modification of the argument in~\cite{wilde_polar_2011}, which in turn came from \cite{mahdavifar_achieving_2011}. First observe that following three sets of channels are disjoint: the 
doubly bad channels $\mathcal B$, the amplitude-good channels to $R$, $\mathcal G_N(\W_R,\beta)$, and the phase-good channels to Bob, $\mathcal G_N(\W_P,\beta)$. Clearly the first and last are disjoint by the definition of $\mathcal B$. The first two are disjoint by the degradability condition: any channel amplitude-bad for Bob must also be amplitude-bad for $R$. Formally, the output fidelity can only go down under the degrading map; see~\cite[Lemma 3]{wilde_polar_2011}. Finally, that the second two are disjoint follows
from the fidelity uncertainty relations in Lemma~\ref{lem:channelfidelityunc}. Setting $\W_1=\W_{P,N}^{(i)}$ and $\W_2=\W_{R,N}^{(i)}$ therein, the lemma states that the fidelities of $\W_{P,N}^{(i)}$ and $\W_{R,N}^{(i)}$ cannot both be small:
\begin{align}
\label{eq:fidelity-uncertainty-relations}
2\cdot2^{-N^{\beta}}\geq2\cdot{F(\W_{P,N}^{\left(  i\right)  })}%
\geq1-{F(\W_{R,N}^{\left(  i\right)  })}.
\end{align}
This implies 
\begin{align}
F(\W_{R,N}^{(i)})\geq 1-2\cdot 2^{-N^\beta},
\end{align}
whenever $2^{-N^{\beta}}\geq{F(W_{P,N}^{\left(  i\right)  })}$.
Thus, all of the channels that are phase-good for Bob are
amplitude-\textquotedblleft very bad\textquotedblright\ for $R$. Therefore the
following relation holds for large enough $N$:%
\begin{align}
\mathcal{G}_{N}\left( \W_P,\beta\right)  \cap\mathcal{G}_{N}\left(
\W_{R},\beta\right)  =\emptyset.
\end{align}

Since these three sets are disjoint, the sum of their sizes cannot exceed $N$, the total number of channels: 
\begin{align}
\frac{1}{N}\left(  \left\vert \mathcal{G}_{N}\left(  W_{R},\beta\right)
\right\vert +\left\vert \mathcal{G}_{N}\left( \W_P,\beta\right)  \right\vert
+\left\vert \mathcal{B}\right\vert \right)  \leq1.
\end{align}
Finally, we know from Theorem~\ref{thm:fraction-good} that the rates of the
sets $\mathcal{G}_{N}\left(  W_{R},\beta\right)  $ and $\mathcal{G}_{N}\left(
W_{P},\beta\right)  $ in the asymptotic limit are%
\begin{align*}
\lim_{N\rightarrow\infty}\frac{1}{N}\left\vert \mathcal{G}_{N}\left(
W_{R},\beta\right)  \right\vert  &  =I\left(  Z^A;R\right)_\psi  ,\\
\lim_{N\rightarrow\infty}\frac{1}{N}\left\vert \mathcal{G}_{N}\left(
W_{P},\beta\right)  \right\vert  &  =I\left(  X^A;BC\right)_\psi  =1-I\left(
Z^A;R\right)_\psi  ,
\end{align*}
so that the rate of $\mathcal{B}$ must be zero in the asymptotic limit. \end{IEEEproof}

Note that in the limit $N\rightarrow \infty$, the channels polarize, so that the channels which are
good in phase for Bob are bad in amplitude for $R$, and the ones which are
good in amplitude for $R$ are bad in phase for Bob. This demonstrates that our
quantum polar coding scheme given here is asymptotically equivalent to the
scheme of Wilde and Guha \cite{wilde_polar_2011}\ in the limit of many recursions of the encoding
after the channel polarization effect takes hold.
Thus, this same argument implies that the entanglement
consumption rate for the quantum polar codes in~\cite{wilde_polar_2011}\ vanishes
for general degradable quantum channels because the rate of the phase-good channels to
Bob is a lower bound on the rate of the amplitude-\textquotedblleft very
bad\textquotedblright\ channels to $R$.


%

\subsection{General Condition Based on Erasure Channels}

The binary erasure channel (BEC) plays a special role in the channel synthesis process. Suppose we have an arbitrary channel $\W$ with output fidelity $F_0$. Then, the $i^{\text{th}}$ channel synthesized from the BEC with fidelity $F_0$ is always less reliable than the corresponding $i^{\text{th}}$ channel synthesized from $\W$.  This is due to the fact that, among all channels with a fixed symmetric capacity, the BEC with that capacity has the smallest output fidelity. This was shown for classical channels in~\cite[Proposition 11]{arikan_channel_2009}, and the same argument works in the quantum case using (21) and (22) of~\cite{wilde_polar_2012-1} instead of (34) and (35) of~\cite{arikan_channel_2009}.  Thus, if the $i^{\text{th}}$ synthesized BEC channel is good, then surely the $i^{\text{th}}$ synthesized $\W$ is also. From this observation, \cite{renes_efficient_2012} gave a condition under which entanglement assistance is needed only at a sublinear rate. Here we show that the same condition holds in the more general setting
discussed in this paper.
\begin{theorem}
\label{thm:fredbound}
In the quantum and private coding schemes above, $\lim_{N\rightarrow \infty}\frac1N|\mathcal B|=0$ if 
\begin{align}
F(\W_A)+F(\W_P)\leq 1.
\end{align}
\end{theorem}
We first provide a heuristic proof sketch which clarifies the main idea behind the proof, and then we follow with the full proof. Since the encoder applies the transformation $G_N$ for the amplitude channel but $G_N^T$ for the phase channel, an input~$i$ corresponds to a doubly-bad synthesized channel if the $i^{\text{th}}$ synthesized amplitude channel is bad and the $(N-i)^{\text{th}}$ synthesized phase channel is bad. For a given input or synthesized channel $i$, call $N-i$ the complementary-variable channel. 

Letting $F_i^p$ be the output fidelity of the $i^{\text{th}}$ channel synthesized from the BEC with erasure probability $p$, the proof rests on the fact that
\begin{align}
F_{N-i}^p=1-F_i^{1-p}.
\end{align}
Since $F_i^{p'}\geq F_i^p$ for $p'\geq p$, this relation implies that, for $p<\nicefrac 12$ the complement of a bad channel is a good channel, while for $p>\nicefrac 12$ the complement of a good channel is a bad channel. Note that $p$ itself is the output fidelity of the BEC with erasure probability $p$.

Now suppose the two channels $\W_A$ and $\W_P$ are erasure channels with erasure probabilities $p_A$ and $p_P$, respectively, such that $p_A\leq p_P<\nicefrac 12$. Thus they satisfy the stated constraint. Indeed, we need only check that no doubly bad channels occur for the case $p_A=p_P$, since then they certainly will not occur when one of the erasure probabilities is smaller and some inputs switch from bad to good. Now, if $i$ is a bad input for $\W_P$, the complementary-variable input is good, and therefore $i$ must be a good input to $\W_A$. Similarly, a bad input for $\W_A$ must be a good input to $\W_P$ and so indeed no doubly-bad inputs can occur in this case. On the other hand, this line of argumentation fails when one or other of the erasure probabilities is greater than $\nicefrac 12$.

Finally, since the BEC is the worst case under channel synthesis among all base channels with a given output fidelity, the stated condition holds for all channels.

\begin{IEEEproof}As described in
Section~\ref{sec:polar-background}, one can prove that channel polarization
takes hold by considering the channel splitting and combining process as a
random birth process $\left\{  \W_{n}:n\geq0\right\}  $ (with the channel
choice determined by an iid Bernoulli process $\left\{  B_{n}:n\geq1\right\}  $
and setting $\W_{0}=\W$). One can then consider the induced birth process for the fidelity parameter%
\[
\left\{  F_{n}:n\geq0\right\}  \equiv\{{F\left( \W_{n}\right)  }%
:n\geq0\}.
\]
In~\cite{arikan_rate_2008} it is shown 
 that the following extremal process $\left\{  F_{n}^{\prime}%
:n\geq0\right\}  $ bounds the actual channel process $\left\{  F_{n}%
:n\geq0\right\}  $:%
\[
F_{n+1}^{\prime}=\left\{
\begin{array}
[c]{ccc}%
F_{n}^{\prime2} & \text{if} & B_{n}=0\\
2F_{n}^{\prime}-F_{n}^{\prime2} & \text{if} & B_{n}=1
\end{array}
\right.  ,
\]
a relation which can be written more symmetrically as%
\begin{align}
F_{n+1}^{\prime}  &  =F_{n}^{\prime2}\ \ \ \ \ \ \ \ \ \ \ \text{if\ \ \ }%
B_{n}=0,\nonumber\\
1-F_{n+1}^{\prime}  &  =\left(  1-F_{n}^{\prime}\right)  ^{2}%
\ \ \ \text{if\ \ \ }B_{n}=1. \label{eq:symmetric-form-birth-proc}%
\end{align}
Note that the extremal process is based on the process for a BEC with the same initial fidelity $F_0$. From now on, we make abbreviations such as $\left\{  F_{n}\right\}  =\left\{
F_{n}:n\geq0\right\}  $ in order to simplify the notation. The extremality of the process is based on recursive relations for the synthesized channel fidelities, equations (34) and (35) of~\cite{arikan_channel_2009}. These have been extended to the case the channel has quantum outputs as equations (21) and (22) of~\cite{wilde_polar_2012-1}. Therefore, the results derived in~\cite{arikan_rate_2008} hold in the present setting as well.

In particular, the extremal process above has the nice property~\cite[Observation 4
(ii)]{arikan_rate_2008} that for every realization $\left\{
b_{n}\right\}  $ of the process $\left\{  B_{n}\right\}  $ (and thus for every
realization $\left\{  f_{n}^{\prime}\right\}  $ of $\left\{  F_{n}^{\prime
}\right\}  $) there exists a particular initial threshold value $F_{\text{th}%
}^{\prime}\left(  \left\{  b_{n}\right\}  \right)  $ such that either%
\[
\lim_{n\rightarrow\infty}f_{n}^{\prime}=0\text{ if }F_{0}^{\prime
}<F_{\text{th}}^{\prime}\left(  \left\{  b_{n}\right\}  \right)  ,
\]
or%
\[
\lim_{n\rightarrow\infty}f_{n}^{\prime}=1\text{ if }F_{0}^{\prime}\geq
F_{\text{th}}^{\prime}\left(  \left\{  b_{n}\right\}  \right)  .
\]
(Note that $F_{0}^{\prime}$ is deterministic and is the initial value of the process.)

We can denote the respective fidelity processes for the amplitude and phase
channels in our coding scheme as $\left\{  F_{n}^{A}\right\}  $ and $\left\{
F_{n}^{P}\right\}  $ and the respective random birth processes as $\left\{
B_{n}^{A}\right\}  $ and $\left\{  B_{n}^{P}\right\}  $. Also, let $\left\{
F_{n}^{A\prime}\right\}  $ and $\left\{  F_{n}^{P\prime}\right\}  $ denote the
corresponding extremal processes. The important observation made in~\cite{renes_efficient_2012} is that the process $\left\{  F_{n}^{P}\right\}  $ makes the
opposite choice of channel at each step of the birth process because the phase
encoder is the reverse of the amplitude encoder. That is, it holds for every
$n$ and for every realization $\left\{  b_{n}^{A}\right\}  $ and $\left\{
b_{n}^{P}\right\}  $ that%
\[
b_{n}^{P}=1-b_{n}^{A}.
\]
Thus, we can write $B_{n}^{P}=1-B_{n}^{A}$, so that $B_{n}^{P}$ is completely
determined by $B_{n}^{A}$. The extremal amplitude channel process $\left\{
F_{n}^{A\prime}\right\}  $\ is already of the form in
(\ref{eq:symmetric-form-birth-proc}), and we can consider the extremal phase
process as $\left\{  1-F_{n}^{P\prime}\right\}  $ in order for it to have this
same form. Thus, a realization $\{f_{n}^{A^{\prime}}\}$ of the extremal
amplitude channel process $\left\{  F_{n}^{A\prime}\right\}  $ converges to
one if%
\[
F_{0}^{A\prime}\geq F_{\text{th}}^{\prime}(\{b_{n}^{A}\}),
\]
and a realization $\{1-f_{n}^{P^{\prime}}\}$ of the extremal phase process
$\left\{  1-F_{n}^{P\prime}\right\}  $ converges to zero if%
\[
1-F_{0}^{P\prime}<F_{\text{th}}^{\prime}(\{b_{n}^{A}\}),
\]
implying that $\left\{  f_{n}^{P\prime}\right\}  $ converges to one if%
\[
F_{0}^{P\prime}>1-F_{\text{th}}^{\prime}(\{b_{n}^{A}\}).
\]
Thus, the sum process $\left\{  F_{n}^{A\prime}+F_{n}^{P\prime}\right\}  $
converges to two if%
\begin{align}
F_{0}^{A\prime}+F_{0}^{P\prime} &  \geq F_{\text{th}}^{\prime}(\{b_{n}%
^{A}\})+1-F_{\text{th}}^{\prime}(\{b_{n}^{A}\})\nonumber\\
&  =1.\label{eq:condition-no-secret-bits}%
\end{align}
The above bound is a \textit{universal}, sufficient lower bound for the sum
process to converge to two, that holds regardless of the threshold value
$F_{\text{th}}^{\prime}(\{b_{n}^{A}\})$ for a particular realization
$\{b_{n}^{A}\}$. It follows that a given realization $\left\{  f_{n}^{A}%
+f_{n}^{P}\right\}  $ of the actual sum process $\left\{  F_{n}^{A}+F_{n}%
^{P}\right\}  $ can only converge to two when
(\ref{eq:condition-no-secret-bits}) holds because we set $F_{0}^{A^{\prime}%
}=F_{0}^{A}$ and the extremal process bounds the actual process (note that
some realizations might converge to one or zero as well). If a realization
$\left\{  f_{n}^{A}+f_{n}^{P}\right\}  $ of the sum process $\left\{
F_{n}^{A}+F_{n}^{P}\right\}  $ converges to two, then this implies that the
set $\mathcal{B}$ is non-empty, i.e., the code will require some preshared entanglement or secret key. So, if the condition in the statement of the theorem holds, no
realization of the sum process can ever converge to two, and the code will not
require any secret key bits.
\end{IEEEproof}

The above argument only holds in the asymptotic limit of many recursions of the encoding
such that the channel polarization effect takes hold (where all synthesized channels are
polarized to be completely perfect or useless). That is, the argument does not apply whenever there
is a finite number of recursions---in this case, if the number of recursions is large enough,
then a large fraction of synthesized channels polarize according to some tolerance, but there
is always a small fraction that have not polarized. Thus, we can only conclude
that the above proof applies in the limit of many recursions and that the rate of secret
key consumption vanishes in this limit.

\section{Superactivation}

\label{sec:superactivation}Our quantum polar coding scheme can be adapted to realize
the superactivation effect, in which two zero-capacity
quantum channels can \textit{activate} each other when used jointly, such that
the joint channel has a non-zero quantum capacity \cite{smith_quantum_2008}. Recall that the
channels from~\cite{smith_quantum_2008} are a four-dimensional PPT\ channel and a
four-dimensional 50\% erasure channel. Each of these have zero quantum
capacity, but the joint tensor-product channel has non-zero capacity.\footnote{We
are speaking of \textit{catalytic} superactivation. A
catalytic protocol  uses entanglement assistance, but the
figure of merit is the net rate of quantum communication---the total quantum
communication rate minus the entanglement consumption rate. Note
that the catalytic quantum capacity is equal to zero
if the standard quantum capacity is zero. Thus, the superactivation effect
that we speak of in this section is for the catalytic quantum capacity.}

We now discuss how to realize a quantum polar coding scheme for the joint
channel. Observe that the input space of the joint channel is 16-dimensional
and thus has a decomposition as a tensor product of four qubit-input
spaces:\ $\mathbb{C}^{4}\otimes\mathbb{C}^{4}\simeq\mathbb{C}^{16}%
\simeq\mathbb{C}^{2}\otimes\mathbb{C}^{2}\otimes\mathbb{C}^{2}\otimes
\mathbb{C}^{2}$. Thus, we can exploit a slightly modified version of our qubit
polar coding scheme. Following~\cite{sasoglu_polarization_2009}, the idea is for Alice and Bob to employ a quantum polar
code for each qubit in the tensor factor. 
Let $Z_{1}$, \ldots, $Z_{4}$ denote the amplitude
variables of these qubits and let $X_{1}$, \ldots, $X_{4}$ denote the phase variables. Bob's
decoder is such that he coherently decodes $Z_{1}$, uses it as quantum side
information (QSI) to decode $Z_{2}$, uses both $Z_{1}$ and $Z_{2}$ as QSI
to decode $Z_{3}$, and then uses all of $Z_{1}$, \ldots, $Z_{3}$
to help decode $Z_{4}$. With all of the amplitude variables decoded, Bob then
uses these as QSI to decode $X_{1}$, and continues
successively until he coherently decodes $X_{4}$. At the end he performs
controlled phase gates to recover entanglement established with Alice.

We now calculate the total rate of this scheme. For the first qubit space in
the tensor factor, the channels split up into four types depending on whether
they are good/bad for amplitude/phase. Using the formula~\eqref{eq:net-rate-calc}, the net quantum
data rate for the first tensor factor is equal to
$$
I\left(  Z_{1};B\right)  +I\left(  X_{1};BZ_{1}Z_{2}Z_{3}Z_{4}\right)  -1.
$$
(The formula is slightly different here because Bob decodes the phase variable
$X_{1}$ with all of the amplitude variables as QSI.) For
the second qubit space in the tensor factor, the net quantum data rate is
$$
I\left(  Z_{2};BZ_{1}\right)  +I\left(  X_{2};BZ_{1}Z_{2}Z_{3}Z_{4}%
X_{1}\right)  -1.
$$
We can similarly determine the respective net quantum data rates for the third
and fourth qubit spaces as
\begin{align*}
& I\left(  Z_{3};BZ_{1}Z_{2}\right)  +I\left(  X_{3};BZ_{1}Z_{2}Z_{3}
Z_{4}X_{1}X_{2}\right)  -1, \\
& I\left(  Z_{4};BZ_{1}Z_{2}Z_{3}\right)  +I\left(  X_{4};BZ_{1}Z_{2}
Z_{3}Z_{4}X_{1}X_{2}X_{3}\right)  -1 .
\end{align*}
Summing all these rates together with the chain rule and using the fact that
any two amplitude and/or phase variables are independent whenever $i\neq j$, we
obtain the overall net quantum data rate:
$$
I\left(  Z_{1}Z_{2}Z_{3}Z_{4};B\right)  +I\left(  X_{1}X_{2}X_{3}X_{4}%
;BZ_{1}Z_{2}Z_{3}Z_{4}\right)  -4,
$$
which is equal to the coherent information of the joint channel (as in the proof of Theorem~\ref{thm:qrate}). The fact that our quantum
polar code can achieve the symmetric coherent information rate then proves
that superactivation occurs, given that Smith and Yard already showed that
this rate is non-zero for the channels mentioned above \cite{smith_quantum_2008}.

\section{Conclusion}

\label{sec:conclusion}
We have demonstrated new polar coding schemes for quantum or private communication which achieve high rates for \emph{arbitrary} quantum channels, unlike the constructions in~\cite{wilde_polar_2011,renes_efficient_2012}. The encoding operations are efficient, though currently no efficient algorithm to construct the code itself. (That is, for general quantum channels, no efficient algorithm is known for determining which inputs correspond to good or bad synthesized channels.) For the decoder the situation is somewhat reversed: Given the code construction, the decoder is explicit---it is based on the quantum successive cancellation method of~\cite{wilde_polar_2012-1}---but no efficient implementation is known. 

Finding an efficient code construction algorithm and an efficient successive cancellation decoder are the main questions left open in this work. It would be interesting to determine conditions beyond those of Section~\ref{sec:noassist} under which entanglement or secret-key assistance are not needed or to find an argument ensuring reliability and strong security of the private coding scheme which relies only on the ``amplitude'' properties of the wiretap channel.

\section*{Acknowledgments}
The authors acknowledge helpful discussions with Fr\'ed\'eric~Dupuis, Saikat~Guha, and
Graeme~Smith.

\appendices

\section{Useful Lemmas}
\label{sec:useful}
\begin{lemma}[Renes \& Boileau~\cite{renes_conjectured_2009}]
\label{lem:certainty}
The following uncertainty relation holds
\begin{align}
H(Z^A|R)_\psi+H(X^A|BC)_\psi=1,
\end{align}
where
$$\ket{\psi}^{ABCR}=\sum_z \sqrt{p_z}\ket{z}^A\ket{z}^{C}\ket{\varphi_z}^{BR}.$$
\end{lemma}
\begin{IEEEproof}
Rewriting system $A$ of $\psi$ using the conjugate basis gives
\begin{align}
\ket{\psi}^{ABCR}&=\sum_z\sqrt{p_z}\tfrac1{\sqrt{2}}\sum_x(-1)^{xz} \ket{\widetilde{x}}^A \ket{z}^{C}\ket{\varphi_z}^{BR}\\
&=\tfrac1{\sqrt{2}}\sum_x\ket{\widetilde{x}}^A (Z^x)^C\ket{\psi'}^{CBR},
\end{align}
where $$\ket{\psi^\prime}^{CBR} = \sum_z \sqrt{p_z}\ket{z}^{C}\ket{\varphi_z}^{BR}.$$ To compute $H(X^A|BC)_{\psi}$ first write it as 
\begin{align*}
H(X^A|BC)_{\psi}=H(X^A)_{\psi}+H(BC|X^A)_{\psi}-H(BC)_{\psi}
\end{align*}
whose terms are easier to evaluate. The first is simply $H(X^A)_{\psi}=1$, while the second is just
$$H(BC|X^A)_{\psi} = H({CB})_{\psi'}=H(R)_{\psi'} = H(R)_{\psi}$$ since the two marginal states of $BC$ given $X^A$ are unitarily-related and one of the marginals is $\psi'^{CB}$. Since $\psi'^{CBR}$ is pure, $H({CB})_{\psi'}=H(R)_{\psi'}$ and a quick calculation reveals that
$H(R)_{\psi'} = H(R)_{\psi}$.
Meanwhile, the $BC$ marginal is $\psi^{ BC}=\sum_z p_z \ket{z}\bra{z}^C\otimes\varphi_z^B$, and so 
\begin{align}
H(CB)_{\psi}&=H(Z^A)_\psi+\sum_z p_zH(\varphi_z^B)\\
&=H(Z^A)_{\psi}+\sum_z p_zH(\varphi_z^R)\\
&=H(Z^AR)_{\psi},
\end{align}
using the entropic properties of bipartite pure states. Thus, 
\begin{align}
H(X^A|BC)_{\psi} &= 1+H(R)_\psi-H(Z^AR)_\psi\\
&=1-H(Z^A|R)_\psi.
\end{align}
\end{IEEEproof}

\begin{lemma}
\label{lem:synth-unc}
The synthesized channels $\W_{P,N}^{(i)}$ and $\W_{R,N}^{(i)}$ obey
\begin{align}
I(\W_{P,N}^{\left(  i\right)  })+I(\W_{R,N}^{\left(  i\right)  }) \leq
1.
\end{align}
The same relation holds when replacing $\W_{P,N}^{(i)}$ and $\W_{R,N}^{(i)}$ with
$\overline{\W}_{P,N}^{(i)}$ and $\overline{\W}_{E,N}^{(i)}$, respectively.
\end{lemma}
\begin{IEEEproof}
 Consider $N$ copies of the channel state $\ket{\psi}^{ABCR}$ in \eqref{eq:channelstate} where the $B$ systems are first subjected to the polarization transformation before input to the channel:
\begin{align*}
\ket{\Psi_N}^{A^NB^NC^NR^N} \!\!&= \tfrac{1}{\sqrt{2^N}}\sum_{z^N}\ket{z^N}^{A^N}\!\ket{z^N}^{C^N}\!\ket{\varphi_{z^NG_N}}^{B^NR^N}
\end{align*} 
Then let $\overline{\Psi}_N^i$ be the state after measuring the systems $A_1\cdots A_{i-1}$ in the amplitude basis and the systems $A_{i+1}\cdots A_N$ in the phase basis, indicating the various measurement output systems as 
$Z_1,\dots Z_{i-1}$ and $X_{i+1},\dots,X_N$ so that it is clear these systems become classical.  From $\overline{\Psi}_N^i$ one can simultaneously generate the outputs of the $i^{\text{th}}$ phase channel to Bob, $\W_{P,N}^{(i)}$ and the $i^{\text{th}}$ amplitude channel $\W_{R,N}^{(i)}$ to the reservoir $R$, just as in the simple case of $\ket{\xi}$ in Section~\ref{sec:complementary-channels}. In particular,  $\overline{\Psi}_N^i$ is a tripartite state on
$$A_i|B^NC^NX^N_{i+1}|R^NZ_1^{i-1},$$where the vertical bars indicate the divisions of the parties. Applying the uncertainty principle from~\cite{renes_conjectured_2009} gives
\begin{align}
H(X^{A_i}|B^NC^NX_{i+1}^N)_{\overline{\Psi}_N^i}+H(Z^{A_i}|R^NZ_{1}^{i-1})_{\overline{\Psi}_N^i} \geq 1
\end{align}
Combining this with $H\left(  X^{A_{i}}\right)
+H\left(  Z^{A_{i}}\right)  =2$, which holds because $X^{A_{i}}$ and
$Z^{A_{i}}$ are uniform random bits, yields%
\[
I\left(  X^{A_{i}};B^{N}C^{N}X_{i+1}^{N}\right)_{\overline{\Psi}_N^i} +I\left(  Z^{A_{i}}%
;R^{N}Z_{1}^{i-1}\right)_{\overline{\Psi}_N^i}  \leq 1,
\]
or equivalently,%
\begin{equation}
I(\W_{P,N}^{\left(  i\right)  })+I(\W_{R,N}^{\left(  i\right)  }) \leq
1.\label{eq:polarized-channel-certainty}%
\end{equation}
The same argument applies starting from $N$ copies of the channel state $\ket{\overline\psi}$ from~\eqref{eq:private-channel-state}.
\end{IEEEproof}

\begin{lemma}
\label{lem:channelfidelityunc}
For complementary binary-input channels $\W_1$ and $\W_2$ obeying the uncertainty relation $I(\W_1)+I(\W_2)\leq 1$,
\begin{align}
2F(\W_1)+\phantom{2}F(\W_2) &\geq 1,\qquad \text{and}\\
\phantom{2}F(\W_1)+2F(\W_2)&\geq 1.
\end{align}
\end{lemma}
\begin{IEEEproof}
Start with the following inequality for binary-input channels $\W$~\cite[{Proposition 1}]{wilde_polar_2012-1}:
\begin{align}
I\left(  \W\right)  \geq\log_{2}\left(  \frac{2}{1+{F\left( \W\right)  }%
}\right). \label{eq:arikan-inequality}%
\end{align}
This is equivalent to $F(\W)\geq 2^{1-I(\W)}-1$. Then we have
\begin{align}
F(\W_1)&\geq 2^{1-I(\W_1)}-1\\
&\geq 2^{I(\W_2)}-1\\
&\geq \frac{2}{1+F(\W_2)}-1,
\end{align}
where we used the uncertainty relation in the second step.  Rewriting this, we obtain
\begin{align}
\left(1+F(\W_2)\right)F(\W_1)\geq 2-\left(1+F(\W_2)\right).
\end{align}
Since the fidelity is less than unity, this gives the first inequality. Interchanging the two channels and repeating the argument gives the second. 
\end{IEEEproof}

\section{Classical Wiretap Channels as Quantum Wiretap Channels}

\label{app:cw-as-qw}Suppose that $p\left(  y,z|x\right)  $ is a classical
wiretap channel such that $x$ is the input and $y$ and $z$ are the outputs for
the legitimate receiver and the wiretapper, respectively. Then we can embed
the random variables $X$, $Y$, and $Z$ into quantum systems, so that the
resulting wiretap channel has the following action on an arbitrary input state
$\rho$:%
\begin{equation}
\mathcal{N}_{\text{C}}^{A^{\prime}\rightarrow BE}\left(  \rho\right)
\equiv\sum_{x,y,z}\left\langle x\right\vert \rho\left\vert x\right\rangle
\ p\left(  y,z|x\right)  \left\vert y\right\rangle \left\langle y\right\vert
^{B}\otimes\left\vert z\right\rangle \left\langle z\right\vert ^{E}.
\label{eq:classical-wiretap}%
\end{equation}
The physical interpretation of the above channel is that it first
\textit{measures} the input system in the orthonormal basis $\left\{
\left\vert x\right\rangle \left\langle x\right\vert \right\}  $ (ensuring that
the input is effectively classical) and \textit{prepares} the classical states
$\left\vert y\right\rangle ^{B}$ and $\left\vert z\right\rangle ^{E}$ for Bob
and Eve with probability $p\left(  y,z|x\right)  $. One can check that the
Kraus operators \cite{wilde_classical_2011}\ for this classical channel are%
\[
\left\{  \sqrt{p\left(  y,z|x\right)  }\left(  \left\vert y\right\rangle
^{B}\otimes\left\vert z\right\rangle ^{E}\right)  \left\langle x\right\vert
^{A^{\prime}}\right\}  _{x,y,z}.
\]
Thus, by a standard construction \cite{wilde_classical_2011}, an isometric extension of this
classical wiretap channel acts as follows on a pure state input $\left\vert
\psi\right\rangle $:%
\begin{align*}
&U_{\mathcal{N}_{\text{C}}}^{A^{\prime}\rightarrow BES_{2}}\left\vert
\psi\right\rangle^{A'}\\
&\quad =
\sum_{x,y,z}\sqrt{p\left(  y,z|x\right)  }\braket{x|\psi}^{A'}\ket{y}^B\ket z^E\ket{x,y,z}^{S_2}
\end{align*}
so that tracing over system $S_{2}$ recovers the action of the original
channel in (\ref{eq:classical-wiretap}).

\begin{figure*}
\begin{equation}
\ket{\Psi_{2;u_{\mathcal{X}},u_{\mathcal{Z}}}}=
\frac{1}{\sqrt{2^{\left\vert \mathcal{A}\right\vert +\left\vert \mathcal{B}%
\right\vert +\left\vert \mathcal{X}\right\vert }}}\sum_{u_{\mathcal{A}%
}^{\prime\prime},u_{\mathcal{B}}^{\prime\prime},v_{\mathcal{X}}^{\prime\prime
}}\left(  -1\right)  ^{u_{\mathcal{X}}\cdot v_{\mathcal{X}}^{\prime\prime}%
}\left\vert u_{\mathcal{A}}^{\prime\prime}\right\rangle \left\vert
\varphi_{u_{\mathcal{A}}^{\prime\prime},u_{\mathcal{Z}},v_{\mathcal{X}}%
^{\prime\prime},u_{\mathcal{B}}^{\prime\prime}}\right\rangle ^{B^{N}E^{N}%
}\left\vert u_{\mathcal{A}}^{\prime\prime}\right\rangle \left\vert
v_{\mathcal{X}}^{\prime\prime}\right\rangle \left\vert u_{\mathcal{B}}%
^{\prime\prime}\right\rangle \left\vert u_{\mathcal{B}}^{\prime\prime
}\right\rangle \left\vert u_{\mathcal{Z}}\right\rangle . \label{eq:1st-step-ideal-state}
\end{equation}
\begin{align}
\ket{\Psi_{2;u_{\mathcal{X}},u_{\mathcal{Z}}}^{\rm actual}}=  \frac{1}{\sqrt{2^{\left\vert \mathcal{A}\right\vert +\left\vert
\mathcal{B}\right\vert +\left\vert \mathcal{X}\right\vert }}}\sum
_{\substack{u_{\mathcal{A}},u_{\mathcal{B}},v_{\mathcal{X}},\\u_{\mathcal{A}%
}^{\prime},v_{\mathcal{X}}^{\prime}}}\left(  -1\right)  ^{u_{\mathcal{X}}\cdot
v_{\mathcal{X}}}\left\vert u_{\mathcal{A}}\right\rangle \sqrt{\Lambda
_{u_{\mathcal{A}}^{\prime},v_{\mathcal{X}}^{\prime}}^{\left(  u_{\mathcal{B}%
},u_{\mathcal{Z}}\right)  }}\left\vert \varphi_{u_{\mathcal{A}},u_{\mathcal{Z}%
},v_{\mathcal{X}},u_{\mathcal{B}}}\right\rangle ^{B^{N}E^{N}}\left\vert
u_{\mathcal{A}}^{\prime}\right\rangle \left\vert v_{\mathcal{X}}^{\prime
}\right\rangle \left\vert u_{\mathcal{B}}\right\rangle \left\vert
u_{\mathcal{B}}\right\rangle \left\vert u_{\mathcal{Z}}\right\rangle  \label{eq:1st-step-actual-state}
\end{align}
\begin{align}
&\braket{\Psi_{2;u_{\mathcal{X}},u_{\mathcal{Z}}}|\Psi_{2;u_{\mathcal{X}},u_{\mathcal{Z}}}^{\rm actual}}  =\frac{1}{2^{\left\vert \mathcal{A}\right\vert +\left\vert \mathcal{B}%
\right\vert +\left\vert \mathcal{X}\right\vert }}\sum_{u_{\mathcal{A}%
},u_{\mathcal{B}},v_{\mathcal{X}},v_{\mathcal{X}}^{\prime}}\left(  -1\right)
^{u_{\mathcal{X}}\cdot\left(  v_{\mathcal{X}}^{\prime}+v_{\mathcal{X}}\right)
}\langle\varphi_{u_{\mathcal{A}},u_{\mathcal{Z}},v_{\mathcal{X}}^{\prime
},u_{\mathcal{B}}}|\sqrt{\Lambda_{u_{\mathcal{A}},v_{\mathcal{X}}^{\prime}%
}^{\left(  u_{\mathcal{B}},u_{\mathcal{Z}}\right)  }}|\varphi_{u_{\mathcal{A}%
},u_{\mathcal{Z}},v_{\mathcal{X}},u_{\mathcal{B}}}\rangle^{B^{N}E^{N}} \label{eq:1st-step-overlap}
\end{align}
\end{figure*}
\begin{figure*}
\begin{align}
&  \mathbb{E}_{U_{\mathcal{X}},U_{\mathcal{Z}}}\left\{  \braket{\Psi_{2;U_{\mathcal{X}},U_{\mathcal{Z}}}|\Psi_{2;U_{\mathcal{X}},U_{\mathcal{Z}}}^{\rm actual}}\right\} \label{eq:1st-step-first} \\
&  =\frac{1}{2^{\left\vert \mathcal{A}\right\vert +\left\vert \mathcal{B}%
\right\vert +\left\vert \mathcal{X}\right\vert +\left\vert \mathcal{Z}%
\right\vert }}\frac{1}{2^{\left\vert \mathcal{X}\right\vert }}\sum
_{u_{\mathcal{X}},u_{\mathcal{Z}}}\sum_{u_{\mathcal{A}},u_{\mathcal{B}%
},v_{\mathcal{X}},v_{\mathcal{X}}^{\prime}}\left(  -1\right)  ^{u_{\mathcal{X}%
}\cdot\left(  v_{\mathcal{X}}^{\prime}+v_{\mathcal{X}}\right)  }\langle
\varphi_{u_{\mathcal{A}},u_{\mathcal{Z}},v_{\mathcal{X}}^{\prime},u_{\mathcal{B}%
}}|\sqrt{\Lambda_{u_{\mathcal{A}},v_{\mathcal{X}}^{\prime}}^{\left(
u_{\mathcal{B}},u_{\mathcal{Z}}\right)  }}|\varphi_{u_{\mathcal{A}}%
,u_{\mathcal{Z}},v_{\mathcal{X}},u_{\mathcal{B}}}\rangle^{B^{N}E^{N}}\\
&  =\frac{1}{2^{\left\vert \mathcal{A}\right\vert +\left\vert \mathcal{B}%
\right\vert +\left\vert \mathcal{X}\right\vert +\left\vert \mathcal{Z}%
\right\vert }}\sum_{u_{\mathcal{A}},u_{\mathcal{B}},v_{\mathcal{X}%
},v_{\mathcal{X}}^{\prime},u_{\mathcal{Z}}}\delta_{v_{\mathcal{X}}^{\prime
},v_{\mathcal{X}}}\langle\varphi_{u_{\mathcal{A}},u_{\mathcal{Z}},v_{\mathcal{X}%
}^{\prime},u_{\mathcal{B}}}|\sqrt{\Lambda_{u_{\mathcal{A}},v_{\mathcal{X}%
}^{\prime}}^{\left(  u_{\mathcal{B}},u_{\mathcal{Z}}\right)  }}|\varphi
_{u_{\mathcal{A}},u_{\mathcal{Z}},v_{\mathcal{X}},u_{\mathcal{B}}}%
\rangle^{B^{N}E^{N}}\\
&  =\frac{1}{2^{\left\vert \mathcal{A}\right\vert +\left\vert \mathcal{B}%
\right\vert +\left\vert \mathcal{X}\right\vert +\left\vert \mathcal{Z}%
\right\vert }}\sum_{u_{\mathcal{A}},u_{\mathcal{B}},v_{\mathcal{X}%
},u_{\mathcal{Z}}}\langle\varphi_{u_{\mathcal{A}},u_{\mathcal{Z}},v_{\mathcal{X}%
},u_{\mathcal{B}}}|\sqrt{\Lambda_{u_{\mathcal{A}},v_{\mathcal{X}}}^{\left(
u_{\mathcal{B}},u_{\mathcal{Z}}\right)  }}|\varphi_{u_{\mathcal{A}}%
,u_{\mathcal{Z}},v_{\mathcal{X}},u_{\mathcal{B}}}\rangle^{B^{N}E^{N}}\\
&  \geq\frac{1}{2^{\left\vert \mathcal{A}\right\vert +\left\vert
\mathcal{B}\right\vert +\left\vert \mathcal{X}\right\vert +\left\vert
\mathcal{Z}\right\vert }}\sum_{u_{\mathcal{A}},u_{\mathcal{B}},v_{\mathcal{X}%
},u_{\mathcal{Z}}}\langle\varphi_{u_{\mathcal{A}},u_{\mathcal{Z}},v_{\mathcal{X}%
},u_{\mathcal{B}}}|\Lambda_{u_{\mathcal{A}},v_{\mathcal{X}}}^{\left(
u_{\mathcal{B}},u_{\mathcal{Z}}\right)  }|\varphi_{u_{\mathcal{A}},u_{\mathcal{Z}%
},v_{\mathcal{X}},u_{\mathcal{B}}}\rangle^{B^{N}E^{N}} \label{eq:1st-step-last}\\
&  \geq1-o(2^{-\frac{1}{2}N^{\beta}}),
\end{align}
\end{figure*}

\section{Detailed Error Analysis for the Coherent Amplitude Decoder}

\label{app:err-analysis-amplitude}We provide details of the error analysis in Section~\ref{sec:err-analysis}
for the first decoding step, in which Bob coherently recovers the amplitude information. Given the frozen bits $u_{\mathcal{X}}$ and $u_{\mathcal{Z}}$,
the ideal state after the first
step of the decoder is given in \eqref{eq:ampidealstate} and for convenience, again in \eqref{eq:1st-step-ideal-state}.
Applying the coherent amplitude measurement $V_A$~in \eqref{eq:amp-decoder} actually results in \eqref{eq:1st-step-actual-state}.
Computing their overlap results in \eqref{eq:1st-step-overlap}.
Next, we take the expectation of their overlap with respect to the uniformly random choice of the
frozen bits $u_{\mathcal{X}}$ and $u_{\mathcal{Z}}$. This leads to the steps in
\eqref{eq:1st-step-first}-\eqref{eq:1st-step-last}, which give the desired result. The penultimate inequality is just $\sqrt{\Lambda}\geq \Lambda$ for any $0\leq \Lambda\leq \mathbbm{1}$. 
The last inequality in this sequence follows from the good performance of the quantum
successive cancellation decoder for the cq amplitude channels~\cite[Proposition 4]{wilde_polar_2012-1}.

\begin{figure*}
\begin{equation}
\ket{\Xi^{\rm ideal}_{x_{\mathcal Z},u_{\mathcal X}}}=\frac{1}{\sqrt{2^{\left\vert \mathcal{A}\right\vert +\left\vert \mathcal{B}%
\right\vert }}}\sum_{x_{\mathcal{A}},x_{\mathcal{B}}}\left\vert \widetilde
{x}_{\mathcal{A}}\right\rangle Z^{x_{\mathcal{A}},u_{\mathcal{X}%
},x_{\mathcal{B}},x_{\mathcal{Z}}}U_{\mathcal{N}}U_{\mathcal{E}}^{A^{\prime
N}}\left\vert \Phi_{\mathcal{A},\mathcal{Z},\mathcal{X},\mathcal{B}%
}\right\rangle \left\vert \widetilde{x}_{\mathcal{A}}\right\rangle \left\vert
\widetilde{x}_{\mathcal{Z}}\right\rangle \left\vert \widetilde{u}%
_{\mathcal{X}}\right\rangle \left\vert \widetilde{x}_{\mathcal{B}%
}\right\rangle , \label{eq:2nd-step-ideal-state}
\end{equation}
\begin{align}
\ket{\Xi^{\rm actual}_{x_{\mathcal Z},u_{\mathcal X}}}&=U_{\mathcal E}^{C^N}V_PU_{\mathcal E}^{\dagger C^N}\left(  \frac{1}{\sqrt{2^{\left\vert
\mathcal{A}\right\vert +\left\vert \mathcal{B}\right\vert  }}}\sum
_{x_{\mathcal{A}},x_{\mathcal{B}}}\left\vert \widetilde{x}_{\mathcal{A}%
}\right\rangle Z^{x_{\mathcal{A}},u_{\mathcal{X}},x_{\mathcal{B}%
},x_{\mathcal{Z}}}U_{\mathcal{N}}U_{\mathcal{E}}^{A^{\prime N}}\left\vert
\Phi_{\mathcal{A},\mathcal{Z},\mathcal{X},\mathcal{B}}\right\rangle \left\vert
\widetilde{x}_{\mathcal{B}}\right\rangle \right)\\
&  =\frac{1}{\sqrt{2^{\left\vert \mathcal{A}\right\vert +\left\vert
\mathcal{B}\right\vert+\left\vert \mathcal Z\right\vert }}}\sum_{x_{\mathcal{A}}^{\prime},x_{\mathcal{Z}%
}^{\prime},x_{\mathcal{B}}^{\prime},x_{\mathcal{A}}}\left\vert \widetilde
{x}_{\mathcal{A}}\right\rangle U_{\mathcal{E}}^{C^{N}}\sqrt{\Gamma
_{x_{\mathcal{A}}^{\prime},x_{\mathcal{Z}}^{\prime}}^{\left(  x_{\mathcal{B}%
},u_{\mathcal{X}}\right)  }}U_{\mathcal{E}}^{\dag C^{N}}Z^{x_{\mathcal{A}%
},u_{\mathcal{X}},x_{\mathcal{B}},x_{\mathcal{Z}}}U_{\mathcal{N}%
}U_{\mathcal{E}}^{A^{\prime N}}\left\vert \Phi_{\mathcal{A},\mathcal{Z}%
,\mathcal{X},\mathcal{B}}\right\rangle \ket{\widetilde{x}'_{\mathcal A}}\ket{\widetilde{x}'_{\mathcal Z}}\ket{\widetilde{u}_{\mathcal X}}\ket{\widetilde{x}_{\mathcal B}} . \label{eq:2nd-step-actual-state}
\end{align}
\begin{align}
&\!\!\!\!\braket{\Xi^{\rm ideal}_{x_{\mathcal Z},u_{\mathcal X}}|\Xi^{\rm actual}_{x_{\mathcal Z},u_{\mathcal X}}} \nonumber\\
&  =\frac{1}{2^{\left\vert \mathcal{A}\right\vert +\left\vert \mathcal{B}%
\right\vert }}\sum_{x_{\mathcal{A}},x_{\mathcal{B}}}\left\langle
\Phi_{\mathcal{A},\mathcal{Z},\mathcal{X},\mathcal{B}}\right\vert
U_{\mathcal{E}}^{\dag A^{\prime N}}U_{\mathcal{N}}^{\dag}Z^{-x_{\mathcal{A}%
},-u_{\mathcal{X}},-x_{\mathcal{B}},-x_{\mathcal{Z}}}U_{\mathcal{E}}^{C^{N}%
}\sqrt{\Gamma_{x_{\mathcal{A}},x_{\mathcal{Z}}}^{\left(  x_{\mathcal{B}%
},u_{\mathcal{X}}\right)  }}U_{\mathcal{E}}^{\dag C^{N}}Z^{x_{\mathcal{A}%
},u_{\mathcal{X}},x_{\mathcal{B}},x_{\mathcal{Z}}}U_{\mathcal{N}%
}U_{\mathcal{E}}^{A^{\prime N}}\left\vert \Phi_{\mathcal{A},\mathcal{Z}%
,\mathcal{X},\mathcal{B}}\right\rangle \label{eq:2nd-step-overlap-1}\\
&  \geq\frac{1}{2^{\left\vert \mathcal{A}\right\vert +\left\vert
\mathcal{B}\right\vert }}\sum_{x_{\mathcal{A}},x_{\mathcal{B}}}\left\langle
\Phi_{\mathcal{A},\mathcal{Z},\mathcal{X},\mathcal{B}}\right\vert
U_{\mathcal{E}}^{\dag A^{\prime N}}U_{\mathcal{N}}^{\dag}Z^{-x_{\mathcal{A}%
},-u_{\mathcal{X}},-x_{\mathcal{B}},-x_{\mathcal{Z}}}U_{\mathcal{E}}^{C^{N}%
}\Gamma_{x_{\mathcal{A}},x_{\mathcal{Z}}}^{\left(  x_{\mathcal{B}%
},u_{\mathcal{X}}\right)  }U_{\mathcal{E}}^{\dag C^{N}}Z^{x_{\mathcal{A}%
},u_{\mathcal{X}},x_{\mathcal{B}},x_{\mathcal{Z}}}U_{\mathcal{N}%
}U_{\mathcal{E}}^{A^{\prime N}}\left\vert \Phi_{\mathcal{A},\mathcal{Z}%
,\mathcal{X},\mathcal{B}}\right\rangle \label{eq:2nd-step-overlap-last}
\end{align}
\end{figure*}
\begin{figure*}
\begin{align}
&  \mathbb{E}_{U_{\mathcal{X}},X_{\mathcal{Z}}}\left\{ \braket{\Xi^{\rm ideal}_{X_{\mathcal Z},U_{\mathcal X}}|\Xi^{\rm actual}_{X_{\mathcal Z},U_{\mathcal X}}} \right\}  \label{eq:2nd-step-analysis-first} \\
&  =\frac{1}{2^{\left\vert \mathcal{A}\right\vert +\left\vert \mathcal{B}%
\right\vert +\left\vert \mathcal{X}\right\vert +\left\vert \mathcal{Z}%
\right\vert }}\sum_{x_{\mathcal{A}},x_{\mathcal{B}},u_{\mathcal{X}%
},x_{\mathcal{Z}}}\left\langle \Phi_{\mathcal{A},\mathcal{Z},\mathcal{X}%
,\mathcal{B}}\right\vert U_{\mathcal{E}}^{\dag A^{\prime N}}U_{\mathcal{N}%
}^{\dag}Z^{-x_{\mathcal{A}},-u_{\mathcal{X}},-x_{\mathcal{B}},-x_{\mathcal{Z}%
}}U_{\mathcal{E}}^{C^{N}}\Gamma_{x_{\mathcal{A}},x_{\mathcal{Z}}}^{\left(
x_{\mathcal{B}},u_{\mathcal{X}}\right)  }U_{\mathcal{E}}^{\dag C^{N}%
}Z^{x_{\mathcal{A}},u_{\mathcal{X}},x_{\mathcal{B}},x_{\mathcal{Z}}%
}U_{\mathcal{N}}U_{\mathcal{E}}^{A^{\prime N}}\left\vert \Phi_{\mathcal{A}%
,\mathcal{Z},\mathcal{X},\mathcal{B}}\right\rangle \\
&  \geq1-o(2^{-\frac{1}{2}N^{\beta}}), \label{eq:2nd-step-analysis-last}
\end{align}
\end{figure*}

\section{Detailed Error Analysis for the Coherent Phase Decoder}
\label{app:err-analysis-phase}
We provide details of the error analysis in Section~\ref{sec:err-analysis}
for the second decoding step, in which Bob coherently recovers the phase information.
We can prove that the phase decoder works well with a uniformly
random choice of the bits $u_{\mathcal{X}}$ and $u_{\mathcal{Z}}$. Observe
that a uniformly random choice of the bits $u_{\mathcal{Z}}$ induces a uniform
distribution of the bits $x_{\mathcal{Z}}$. Let us fix a value of $x_{\mathcal Z}$. Then, a similar error analysis as
in Appendix~\ref{app:err-analysis-amplitude}
then works for this case. The ideal state resulting from the second decoding step
is given in \eqref{eq:2nd-step-ideal-state}. This state is the same as in~\eqref{eq:ideal-phase-state} with a fixed, but randomly-chosen value of $x_{\mathcal Z}$. 
The actual state from using the coherent decoder $V_P$ in~\ref{eq:phasedecoder} is given in \eqref{eq:2nd-step-actual-state}.
The overlap between the above two states is analyzed
in \eqref{eq:2nd-step-overlap-1}-\eqref{eq:2nd-step-overlap-last}.
Taking the expectation of this term over a uniformly random choice of
$u_{\mathcal{X}}$ and $u_{\mathcal{Z}}$ (which implies a uniformly random
choice of $x_{\mathcal{Z}}$) leads to the steps in
\eqref{eq:2nd-step-analysis-first}-\eqref{eq:2nd-step-analysis-last}.
The last inequality of this sequence again follows from the performance of the quantum
successive cancellation decoder for the phase channels.

\bibliographystyle{IEEEtran}
\bibliography{quantumpolar}

\end{document}